\documentclass[iop,apj]{emulateapj}
\slugcomment{Accepted for publication in The Astrophysical Journal}

\usepackage[]{amsmath}
\usepackage[dvipsnames]{xcolor}
\usepackage{bm}
\usepackage{ulem}
\shorttitle{Star formation histories of massive dead galaxies at $z\sim2$}
\shortauthors{Morishita et al.}

\usepackage{color}			      
\definecolor{midgray}{gray}{0.4}		
\definecolor{orange}{rgb}{1,0.5,0}    
\usepackage{mathptmx}
\usepackage{times}

\definecolor{ao}{rgb}{0.0, 0.0, 1.0}
\usepackage[colorlinks=true, citecolor=ao, filecolor=ao, linkcolor=ao, urlcolor=ao]{hyperref} 
\usepackage{scrextend}
\deffootnote[0.5em]{0em}{1em}{\textsuperscript{\thefootnotemark}\,}

\usepackage{etoolbox}
\makeatletter
\patchcmd{\NAT@citex}
  {\@citea\NAT@hyper@{\NAT@nmfmt{\NAT@nm}\NAT@date}}
  {\@citea\NAT@nmfmt{\NAT@nm}\NAT@hyper@{\NAT@date}}
  {}
  {}
\patchcmd{\NAT@citex}
  {\@citea\NAT@hyper@{%
     \NAT@nmfmt{\NAT@nm}%
     \hyper@natlinkbreak{\NAT@aysep\NAT@spacechar}{\@citeb\@extra@b@citeb}%
     \NAT@date}}
  {\@citea\NAT@nmfmt{\NAT@nm}%
   \NAT@aysep\NAT@spacechar%
   \NAT@hyper@{\NAT@date}}
  {}
  {}
\patchcmd{\NAT@citex}
  {\@citea\NAT@hyper@{%
     \NAT@nmfmt{\NAT@nm}%
     \hyper@natlinkbreak{\NAT@spacechar\NAT@@open\if*#1*\else#1\NAT@spacechar\fi}%
       {\@citeb\@extra@b@citeb}%
     \NAT@date}}
  {\@citea\NAT@nmfmt{\NAT@nm}%
   \NAT@spacechar\NAT@@open\if*#1*\else#1\NAT@spacechar\fi%
   \NAT@hyper@{\NAT@date}}
  {}
  {}
\makeatother

\usepackage{booktabs}

\newcommand{\myemail}{tmorishita@stsci.edu}
\newcommand{\simgt}{\,\rlap{\lower 3.5 pt \hbox{$\mathchar \sim$}} \raise
1pt \hbox {$>$}\,}
\newcommand{\simlt}{\,\rlap{\lower 3.5 pt \hbox{$\mathchar \sim$}} \raise
1pt \hbox {$<$}\,}
\newcommand{\logm}{\log M_*/\Msun}
\newcommand{\Msun}{M_{\odot}}

\newcommand{\logZ}{\log Z_*/Z_\odot}
\newcommand{\chinu}{\chi^2/\,\nu}

\newcommand{\kms}{{\rm km~s^{-1}}}
\newcommand{\oii}{[\textrm{O}~\textsc{ii}]}
\newcommand{\oiii}{[\textrm{O}~\textsc{iii}]}
\newcommand{\nii}{[\textrm{N}~\textsc{ii}]}

\newcommand{\hd}{\textrm{H}\textsc{$\delta$}}
\newcommand{\hg}{\textrm{H}\textsc{$\gamma$}}
\newcommand{\hb}{\textrm{H}\textsc{$\beta$}}
\newcommand{\ha}{\textrm{H}\textsc{$\alpha$}}

\newcommand{\Dn}{${\rm D_n 4000}$}

\def\ngdn{51} 
\def\ngds{73} 
\def\nmacs{17} 

\def\ngdnf{12} 
\def\ngdsf{10} 
\def\nmacsf{2} 

\def\Ns{24} 

\newcommand{\hst}{{\it HST}}

\newcommand{\spit}{{\it Spitzer}}

\newcommand{\Z}{{\rm [Z/H]}}
\newcommand{\FH}{{\rm [Fe/H]}}

\newcommand{\alp}{{\rm [$\alpha$/Fe]}}

\newcommand{\sext}{{SExtractor}}

\newcommand{\gsf}{{\ttfamily gsf}}

\newcommand{\affilA}{Space Telescope Science Institute, 3700 San Martin Drive, Baltimore, MD 21218, USA}
\newcommand{\affilB}{The Observatories of the Carnegie Institution for Science, 813 Santa Barbara St., Pasadena, CA 91101, USA}
\newcommand{\affilC}{Department of Physics and Astronomy, UCLA, 430 Portola Plaza, Los Angeles, CA 90095-1547, USA}
\newcommand{\affilD}{Cosmic Dawn Centre, University of Copenhagen, Blegdamsvej 17, 2100 Copenhagen, Denmark}
\newcommand{\affilE}{University of California Davis, 1 Shields Avenue, Davis, CA 95616, USA}
\newcommand{\affilF}{School of Physics and Astronomy, University of Minnesota,116 Church Street SE, Minneapolis, MN 55455, USA}
\newcommand{\affilG}{School of Physics, Tin Alley, University of Melbourne VIC 3010, Australia}
\newcommand{\affilH}{ARC Centre of Excellence for All-Sky Astrophysics in 3 Dimensions, Australia}
\newcommand{\affilI}{INAF -- Osservatorio Astronomico di Padova, Vicolo Osservatorio 5, IT-35122, Padova, Italy}

\begin{document}
\title{
Massive Dead Galaxies at {\bm $z\sim2$} with \hst\ Grism Spectroscopy. I. Star Formation Histories and Metallicity Enrichment
}

\author{
T.~Morishita\altaffilmark{1},
L.~E.~Abramson\altaffilmark{2}, 
T.~Treu\altaffilmark{3},
G.~B.~Brammer\altaffilmark{1,4},
T.~Jones\altaffilmark{5},
P.~Kelly\altaffilmark{6},
M.~Stiavelli\altaffilmark{1},
M.~Trenti\altaffilmark{7,8},
B.~Vulcani\altaffilmark{9},
X.~Wang\altaffilmark{3}
}
\affil{$^1$\affilA; \href{mailto:\myemail}{\myemail}}
\affil{$^2$\affilB}
\affil{$^3$\affilC}
\affil{$^4$\affilD}
\affil{$^5$\affilE}
\affil{$^6$\affilF}
\affil{$^7$\affilG}
\affil{$^8$\affilH}
\affil{$^9$\affilI}
\begin{abstract}
Observations have revealed massive ($\logm\simgt11$) galaxies that were already dead when the universe was only $\sim2$\,Gyr. Given the short time before these galaxies were quenched, their past histories and quenching mechanism(s) are of particular interest. In this paper, we study star formation histories (SFHs) of 24 massive galaxies at $1.6<z<2.5$. A deep slitless spectroscopy + imaging data set collected from multiple {\it Hubble Space Telescope} surveys allows robust determination of their spectral energy distributions and SFHs with no functional assumption on their forms. We find that most of our massive galaxies had formed $> 50\%$ of their extant masses by $\sim1.5$\,Gyr before the time of observed redshifts, with a trend where more massive galaxies form earlier. Their stellar-phase metallicities are already compatible with those of local early-type galaxies, with a median value of $\logZ = 0.25$ and scatter of $\sim0.15$\,dex. In combination with the reconstructed SFHs, we reveal {their rapid metallicity evolution} from $z \sim 5.5$ to $\sim 2.2$ at a rate of $\sim0.2$\,dex\,Gyr$^{-1}$ in $\logZ$. Interestingly, the inferred stellar-phase metallicities are, when compared at half-mass time, $\sim0.25$\,dex higher than observed gas-phase metallicities of star forming galaxies. {While systematic uncertainties remain, this may imply that these quenched galaxies have continued low-level star formation, rather than abruptly terminating their star formation activity, and kept enhancing their metallicity until recently.} 
\end{abstract}



\section{Introduction}\label{sec:intro}

In the local universe, early-type galaxies dominate the massive end of the galaxy mass function, $\logm\simgt11.5$ \citep[][]{cole01,bell03}. Those galaxies consist of old and chemically enriched stellar populations, indicating that most of their star formation activities ended $\simgt10$\,Gyr \citep{kauffmann03,thomas03,thomas10,gallazzi05,treu05}. In fact, observations have revealed that some galaxies are already massive and passively evolving at $z\gtrsim2$ \citep[][]{cimatti04,daddi05,vandokkum08,kriek09,straatman14, belli14,marsan15,glazebrook17}. Given the short time since the big bang and their stellar mass, their earlier star formation must be extremely intense, followed by a {rapid} cessation of their star formation activity, which we here refer to as quenching.\footnote{The term may refer to different phenomena in different contexts. For example, one may also refer to keeping star formation at very low levels after an initial decline, or the (rapid) decline of SF itself \citep[see][for a recent review]{man18}. We use the term to describe any decline of galaxy star formation activity regardless of speed.}

However, these episodes still remain observationally indirect. What were their star formation histories (SFHs) like? How and why did they stop forming stars, especially at the peak time of the cosmic star formation? Are they already enriched in metallicity as local counterparts, or do any post-quenching processes play key roles over the following 10\,Gyr? These are the central questions we aim to answer in this series of papers. In this first paper, we focus on their SFHs. 

A number of studies have investigated SFHs of massive galaxies in different approaches. For example, observations of high-redshift galaxies provide an analogy to their past properties, especially when they were actively forming stars. While sufficient valuable information can be obtained from high-$z$ populations \citep[e.g., star formation rate (SFR), number density, metallicity; e.g.,][]{hamann99, tacconi08, toft14}, it is limited by its rather indirect aspect, where connecting different objects at different epochs may introduce systematic uncertainties \citep[e.g.,][]{wellons15, torrey17}.

Another approach is based on the archeological information of local galaxies, or fossil record \citep{thomas03,thomas10, heavens04,mcdermid15}. Detailed information about their stellar population (e.g., age and chemical abundances) provides their past histories, characterizing their formation redshift to $\simgt2$. This information from the local galaxies is, however, limited up to several Gyr with current observing facilities \citep[e.g.,][see also \citealt{conroy13}]{worthey94}, which is short for exploring the SFHs of galaxies that are already dead at $z\sim2$.

To explore evolution histories in the earlier epoch, we need a method that combines some of the virtues of both approaches, that is, the archeological study of high-$z$ galaxies. For example, \citet{kelson14} performed spectral energy distribution (SED) modeling of galaxies at $z\sim1$ by using low-resolution spectra and broadband photometry and reconstructed their SFHs back to $z\sim1.5$ \citep[also][]{dressler16,dressler18}. \citet[][]{chauke18} recently attempted a similar approach to galaxies at $z\simlt1$ but with higher-resolution spectra taken with a ground-based spectrograph and successfully revealed their formation histories back to $z\sim2.5$. In the current study, we target galaxies at higher redshift, aiming at earlier evolution histories up to their formation redshift from the fossil record obtained with the low-spectral resolution yet high-sensitivity Hubble Space Telescope (\hst) spectrophotometric data set.\footnote{\citet{belli19} recently presented star formation histories of $1.5<z<2.5$ galaxies reconstructed with ground-based spectroscopic data. None of their sample galaxies overlaps with ours in this study.}

Stellar metallicity is another key parameter that provides further details of physical mechanisms. In particular, since both the cosmic metallicity and individual gas-phase metallicity are still pristine at these redshifts \citep[e.g.,][]{erb06,maiolino08,lehner16}, the enrichment process within such massive systems has to be substantial to explain the observed solar/supersolar metallicity of lower-redshift galaxies \citep{onodera12,gallazzi14,choi14,lonoce15}, whereas the process is highly dependent on SFHs \citep[e.g.,][]{peng15}. 

With such demands, we here improve our previous methodology of SED modeling, which is free from functional forms of SFHs, by increasing the flexibility in metallicity. We collect \Ns\ massive quenched galaxies at $z\sim2$ that have deep WFC3/G102 and G141 grism spectra coverage in their rest-frame 4000\,\AA. The combination of grism spectra and wide broadband photometry ($0.2$--$8.0\,\mu{\rm m}$ by \hst\ and \spit) provides a unique opportunity to constrain not only age but also metallicity from the entire SED shape.

We proceed as follows. In Section~\ref{sec:data}, we describe the data used in this study and their reduction process. In Section~\ref{sec:sed}, we introduce our method for the SED modeling. In Section~\ref{sec:result}, we show the results. We discuss our results and interpretation in Section \ref{sec:disc} and close in Section~\ref{sec:sum}. Further details, including intensive simulation tests of SED modeling and comparison with functional SFHs, are also presented in the Appendices. Throughout the text, magnitudes are quoted in the AB system \citep{oke83,fukugita96}; $\Omega_m=0.27$, $\Omega_\Lambda=0.73$, $H_0=72\,\kms\, {\rm Mpc}^{-1}$ for the cosmological parameters; and $Z_\odot=0.0142$ \citep{asplund09} for the solar metallicity.


\section{Data}
\label{sec:data}
To achieve our goal of constraining galaxy SFHs, it is essential to cover the wavelength range surrounding 4000\,\AA, where the spectral features are most age-sensitive, with sufficiently deep spectra and rest-frame NUV-optical-NIR wavelength with broad band photometry. Therefore, we limit ourselves to only fields where deep \hst\ grism data are available, which are MACS1149.6+2223 (hereafter M1149) and the GOODS-North/South (GDN/GDS). To collect the initial photometric sample galaxies, we use publicly available photometric catalogs. We use \hst/WFC3 G102 and G141 grism data (that covers $\lambda_{\rm obs}\sim8000$\,\AA\ to 17000\,\AA) taken in various surveys in these fields.

\subsection{Initial Photometric Sample}
M1149 is a sightline of a massive cluster of galaxies at $z=0.544$. The data were taken in CLASH \citep{postman12}, Hubble Frontier Fields \citep[][]{lotz17}, GLASS \citep{schmidt14, treu15}, and the SN Refsdal follow-up campaigns \citep{kelly15,kelly16}. The combination of its gravitational magnification power and those very deep observations provides a unique opportunity among other clusters of GLASS. We use the photometric catalog used in \citet{morishita17, morishita18}, which consists of \hst\ photometry taken in all \hst\ surveys above, as well as \spit\ IRAC ($3.6+4.5\,\mu$m; PIs: T. Soifer and P. Capak) and ground-based $K_S$-band imaging \citep[][]{brammer16} and publicly available spectroscopic redshifts by GLASS \citep{schmidt14}.

For GDN and GDS, we use the publicly available catalog by 3DHST \citep{skelton14}. The photometric catalog consists of fluxes, spectroscopic/photometric redshifts, rest-frame colors, and stellar mass, based on data taken in CANDELS and 3DHST \citep{grogin11,koekemoer11,vandokkum13b,skelton14,momcheva16}, as well as photometric fluxes obtained in ground-based surveys. We use \hst\ photometry for optical/near-IR range, ground-based $K_S$-band photometry, and \spit\ IRAC photometry in this study. Other ground-based fluxes listed in the 3DHST catalog, most of which are in the optical range, are not used, as the deep \hst\ photometry is sufficient to constrain optical SEDs.

Broadband fluxes of \hst\ are measured in a fixed aperture ($r=0.7$\,\arcsec) and then scaled to the total flux by multiplying $C=f_{\rm AUTO}/f_{\rm aper}$, where $f_{\rm AUTO}$ is AUTO flux of \sext\ \citep{bertin96}, as in \citet{skelton14} and \citet{morishita17}.

\begin{figure*}
\centering
	\includegraphics[width=0.8\textwidth]{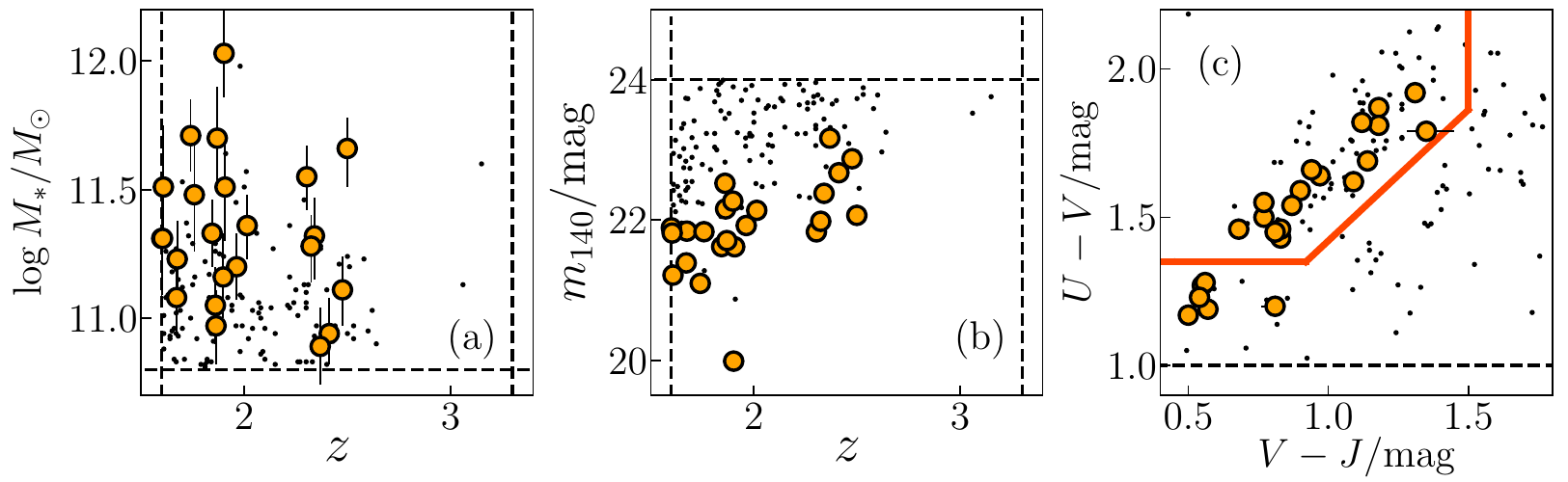}
	\caption{Physical properties of the initial photometric sample galaxies (dots) and final \Ns\ galaxies (orange circles). The criteria used for the initial photometric selection are shown with dashed lines.
	(a) Stellar-mass distribution as a function of redshift.
	(b) The F140W magnitude distribution as a function of redshift.
	(c) The UVJ-color diagram. Our sample consists of both quenched (left-top region bounded by solid lines) and quenching (those $U-V>1$ but outside the boundary) galaxies. 
	}
\label{fig:mz}
\end{figure*}

From these photometric catalogs, we choose those satisfy $m_{140} < 24$, $1.6<z<3.3$, $\logm>10.8$. We also apply $(U-V)_{\rm rest} > 1$\,mag to select quiescent galaxies. By setting slightly bluer color for $U-V$ than literature ($\sim1.4$\,mag), our sample also contains quenching galaxies. With the criteria, we found \nmacs, \ngdn, and \ngds\ galaxies in M1149, GDN, and GDS, respectively, as an initial photometric sample (Figure~\ref{fig:mz}).

\subsection{\hst\ Grism Spectrum}
\label{sec:spec}

In M1149, the grism data were taken through GLASS \citep{schmidt14, treu15}.
GLASS is a spectroscopic survey with \hst/WFC3 G102 and G141 grisms (10 and 4\,orbits, respectively).
In addition, we supplement with the follow-up HST-GO/DDT campaign (Proposal ID 14041, PI: P. Kelly) of the multiply imaged supernova, SN Refsdal \citep{kelly15,kelly16}, which adds another 30\,orbits of G141 data to the original GLASS observation.

In GDN/GDS, we retrieve the public data through MAST. In addition to the 3DHST data \citep{vandokkum13, momcheva16} that cover entire CANDELS's GDN/GDS fields, we add those taken in FIGS \citep[13779; PI: S. Malhotra][]{pirzkal17}, CLEAR \citep[14227; PI: C. Papovich, also][]{carpenter18}, and other follow-ups (12099\&12461; PI: A. Riess, 12190; A. Koekemoer, 13420; PI: G. Barro, 13871; PI: P. Oesch).

We extract 1D spectra from all fields in a consistent way, by using the latest version of Grizli \citep{brammer18}. During the extraction, the code automatically models neighboring objects, which are flagged in pre-provided \sext\ segmentation maps, and produces clean spectra for a target galaxy. The clean, optimal extracted spectra from each position angle (PA) is then stacked in a refined wavelength grid of $45$\,\AA/\,pixel. The pixel scale is slightly finer than the Nyquist sampling of G141 grism, since we have many sampling over different orbits, each of which slightly shifts in the dispersion direction. Each spectrum is convolved with the image of the source to match the morphological difference in different PAs. 

For the aperture correction of broadband photometry, we match the pseudo-broadband flux extracted from grism spectra by convolving with the corresponding filters (F140W/F105W) to the observed broadband flux (Section~\ref{ssec:mcmc}).

In addition to the random uncertainty in flux, we also estimate the uncertainty associated with the stacking of different PAs by following \citet{onodera15}, and we integrate this to the random noise in quadrature for conservative estimates. The uncertainty accounts of $\sim20\%$ of the random uncertainty. {Signal-to-noise rations (S/Ns) of the final 1D spectrum range up to $\sim50$. Median values of each spectral element are ${\rm S/N}\sim18$ at $4200$--$5000$\,\AA\ and $\sim4$ at $3400$--$3800$\,\AA\ (Table~\ref{tab:sed}).}

\subsection{Additional Photometric Data}\label{ssec:addphot}
In addition to photometric fluxes collected in 3DHST, we add WFC3/{\it UVIS} photometry. The rest frame UV coverage is important to constrain SEDs with/without the UV upturn, which depends on metallicity \citep[e.g.,][]{yi97, treu05b}. We use WFC3/{\it UVIS} images from HDUV legacy survey \citep{oesch18b}, that cover parts of GDN/GDS fields with F275W and F336W filters. The {\it UVIS} images in GDS consist of the previous data taken in the UDF by the UVUDF team \citep{teplitz13}. We run \sext\ on the public imaging data to conduct photometry and use the flux measured in a fixed aperture of 0.\!\arcsec72 diameter.

\begin{figure*}
\centering
	\includegraphics[width=0.9\textwidth]{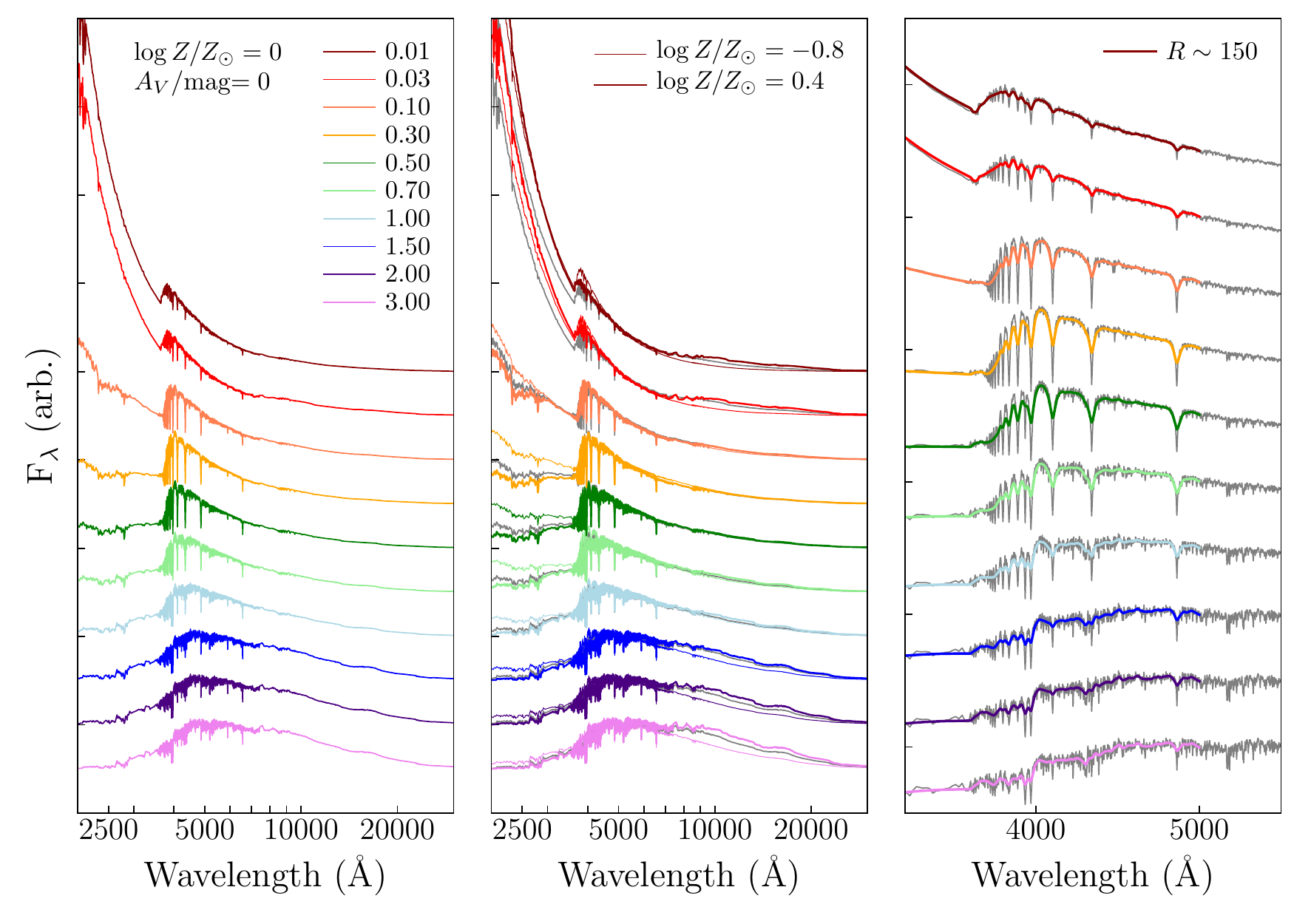}
	\caption{
	{\bf Left:} Original spectral templates used in fitting. Only those with $\logZ=0$ without dust attenuation ($A_V=0$) are shown here, with arbitral shifts in flux. 
	{\bf Left:} Same as left panel (gray solid lines) but also showing templates with different metallicities for comparison (thin lines for $\logZ=-0.8$; thick lines for $\logZ=0.4$). Templates are normalized at $5000$\,\AA.
	{\bf Right:} Same as left panel (gray solid lines) but also showing spectra that are degraded to match the observation, that accounts for the G102/G141 resolutions and source morphology ($R\sim150$; colored lines).
	}
\label{fig:demo}
\end{figure*}

\section{Spectrophotometric SED fitting}\label{sec:sed}

\subsection{Basic Templates}\label{ssec:template}
Our SED fitting method (Grism SED Fitter, or \gsf; Morishita, in prep.) is based on the canonical template fitting, where the best-fit parameters are determined by minimizing the residuals of observed and model SEDs. One major difference from most of other works is the way we construct the model templates. SED templates are often constructed with a functional form for SFHs, such as the exponential declining model, $\psi \propto A \exp[-(t-T_0)/\tau]$, where $A$, $T_0$, and $\tau$ are free parameters. However, it is known that such a simplification may not represent real galaxy SFHs by observations \citep[e.g.,][]{pacifici16, iyer17} and simulations \citep[e.g.,][]{diemer17}. As such, we here avoid any functional forms and adopt an alternative method to generate model templates. The core of the method is to find the best combination of amplitudes, $\{a_{i}\}$, {for a set of composite stellar population (CSP) templates of different ages}, $\{t_{i}\}$, that matches the data, as previously performed by \citet{morishita18}.
{This type of SED modeling has been used in previous studies \citep{heavens04,cidfernandes05,panter07,tojeiro07,kelson14,dressler18}, some of which demonstrated its strength and validity with intensive simulation tests.}

To generate the template with different parameters, we use the flexible stellar population synthesis code \citep[FSPS;][]{conroy09fsps, conroy10,foreman14} to generate $i$th templates with ages of $t_{i}$, based on MIST isochrones \citep{choi16} and the MILES stellar library. As found by \citet{morishita18}, different isochrones may return different results, in addition to a systematic difference in assumed metallicities. 

We set the number of age ``pixels'' to 10, with [0.01, 0.03, 0.1, 0.3, 0.5, 0.7, 1.0, 1.5, 2.0, 3.0]\,Gyr (Figure~\ref{fig:demo}), doubling the number from those adopted in \citet{morishita18}. While we set the equal width of template in log normal space ($\sim0.5$) following previous studies \citep[e.g.,][]{cidfernandes05}, we added extra bins at intermediate age, where most of our galaxies are located, to increase the flexibility of SFHs.

The template is generated by assuming a short constant star formation rate within each bin width ($\sim30$\,Myr), rather than a simple stellar population (SSP). The reason we do not adopt the SSP model is that, while it is simple, it is unrealistic for real galaxies. Changing the width of constant star formation in each bin would result in a minor but systematic shift in reconstructed SFHs. {The uncertainty in bin width is considered in calculation of parameters (e.g., age), by randomly fluctuating values within the width.}

We also set metallicity of each age pixel as a free parameter in a range of $\logZ\in[-0.8:0.6]$, as opposed to one global value in \citet{morishita18}. While determination of metallicity at each age pixel (i.e. metallicity histories of individual galaxies) is more challenging (see Appendix A), this gives extra flexibility in fitting templates, and allows reasonable estimate of uncertainty in SFHs (Section~\ref{ssec:Zh}).

It is noted that metallicity sensitive lines (such as Fe and Mg) are not measured at our spectral resolution. Our method rather relies on the entire spectral shape with grism spectra and wide broadband photometry, that spans from NUV, optical (that are sensitive to age), to NIR (to metallicity) wavelength range (see Figure~\ref{fig:demo}).
 
Templates generated with MIST are uniformly set to the solar-scaled abundance (\citealt{asplund09}; i.e. $[\alpha/{\rm Fe}]=0$). It is noted that galaxies at high $z$ may have an $\alpha$-enhanced chemical composition \citep[e.g.,][]{onodera15,kriek16}, as found in local early-type galaxies \citep[e.g.,][]{thomas05,walcher15}. In fact, enhancement of $\alpha$-element has a similar effect as that of iron in UV and NIR continuum slopes \citep[e.g.,][]{vazdekis15}, while our low-resolution spectra cannot capture a detailed difference in each absorption line (i.e. Lick indices), and both abundances are degenerated in our total metallicity measurement, $\logZ$.\footnote{Total metallicity is often inferred with ${\rm[Z/H]} = {\rm[Fe/H]} + A{\rm[\alpha/Fe]}$, where $A\sim0.9$ depending on abundance ratios \citep[e.g.,][]{trager00, jimenez07}.} As such, our total metallicity should remain similar to those with, e.g., $\alpha$-enhanced templates \citep[see also][]{walcher09}.

We assume a \citet{salpeter55} initial mass function and \citet{calzetti00} dust law, where the dust attenuation, $A_V$, is a global parameter that is applied to all age templates equally. {Redshift is set as a free parameter at this step but within the $3\,\sigma$ range estimated in the previous step}. In sum, the fits have $10\times2+1+1=22$ free parameters.

The degree of freedom of our fitting is worth noting. The number of spectral data points for each of our galaxies is $>200$ (with $\sim16$ for broadband photometric data points), where the spectral element is set to $45$\,\AA\ in this study. Considering the correlation due to morphology (which is $\sim104$\,\AA\ for the mean size of our galaxies, $r\sim0.3\arcsec$), and the large number of parameters, our spectra still have $\sim100$ independent data points.

Our updated method here has a few advantages over \citet{morishita18}. First, it is more flexible than an a priori assumption of the SFH, and robust to systematic bias in derived parameters \citep[e.g.,][]{wuyts12}. Second, it is flexible to a complex shape of SFHs, such as those with multiple bursts and sudden declines \citep[e.g.,][]{boquien14}. A SFH with multiple peaks cannot be reproduced by the exponential declining model (see Section~\ref{sec:result}). Third, it is flexible to the metallicity evolution. Methods with functional forms often have a fixed metallicity over the entire history. In our method, each template at a different age has a flexibility in metallicity as a free parameter that also provides metallicity enrichment histories, {though the uncertainty in each age is typically large for most of our data sets in this study (Appendix~A)}.

\subsection{SED parameter exploration}\label{ssec:mcmc}
The combination of templates is controlled by changing each amplitude, $a_i$, as free parameters during the fit. A challenging part is the large number of parameters (over a dozen, compared to $\simlt5$ parameters with functional SFHs), which could be trapped in local minima. To sufficiently, yet efficiently, explore the parameter space, we adopt the Markov chain Monte Carlo (MCMC) method. 

The fitting process of \gsf\ is twofold: (1) initial redshift determination based on visual inspection of absorption lines, and (2) MCMC realization to estimate the probability distribution for all parameters.

First, \gsf\ determines the redshift by fitting the model templates to the observed grism spectra. At this point, it only generates model templates in the wavelength range of grism spectra, to minimize the computational cost. The templates are convolved to the resolution of the spectra with a Moffat function derived from the observed source morphology for spectra. It searches the best-fit redshift by minimizing $\chi^2$, as well as visual inspection, to avoid catastrophic errors. During the visual inspection, we rely on the major absorption lines in the observed range (i.e. H$\delta$, H$\gamma$, H$\beta$), and thus those without clear absorption features (i.e. low S/N) are discarded here. At this step, \gsf\ also determines the scale of the G102/G141 spectra so that each matches to the broadband photometry in F105W and F140W at a given template. The added scale for our sample is small ($\simlt 10\%$) thanks to the accurate sky-background estimation in Grizli.

Then, \gsf\ generates a template library at the redshift determined in the previous step and fits SEDs at the entire wavelength range. \gsf\ fits the observed spectra and broadband photometry simultaneously by using {\it emcee} \citep[][]{foreman13} as in \citet{morishita18}. Redshift is also explored at this step by shifting and refining the template wavelength grid at a proposed redshift of each MCMC step. Emission lines, if detected, are masked during the fit. Those lines are modeled with a gaussian function after subtracting the best-fit SED template to estimate the line flux and equivalent width (EW; Section~\ref{ssec:spectra}).

We set the number of walkers to 100 and the number of realization, $N_{\rm mc}$, to $10^5$. We adopt an uniform prior for each parameter over the parameter ranges, $a_i \in[0:1000]$, $\logZ\in[-0.8:0.6]$, and $A_V/$mag$\in[0:4]$. {The effect by the amplitude prior (i.e. SFH; see Figure~\ref{fig:simSFH}) should be minimal due to the wide constrain range, as seen in the simulation in Appendix~A.} While some of previous studies set a prior in metallicity histories from the local mass metallicity relation \citep[e.g.,][]{pacifici16,leja18b} (i.e. increasing metallicity as a function of time), we find that our result reproduces this behavior without such priors. However, the age-metallicity degeneracy could also mimic this trend. Our test with a mock data set (Appendix~A) revealed, while the trend is fairly reproduced, nonnegligible scatters in derived metallicity at each age pixel, and that metallicity histories of individual galaxies remain less promising (see also below). 

During the fit, we let {\it emcee} run with the parallel tempering sampling, with $n_{\rm temp}=5$. With this, {\it emcee} samples the parameter spaces but with $n_{\rm temp}$ samplers in parallel. Each sampler has a different value for the temperature parameter in a Metropolis--Hastings step, i.e. a higher temperature makes a larger step in a parameter space. With this, the sampler suffers less from the local minima. We note, however, that a sufficient number of MCMC realizations (which is inferred from a simulation test for our case) is necessary to estimate reliable uncertainties. Otherwise, derived uncertainties would become inappropriately small, with possible biases in the best-fit values \citep[see][]{fernandes18}.

The first half realization of the sampled chain is discarded to avoid biased results from initial input values. We take the 50th and 16th/84th percentiles of marginalized distributions as the best-fit and uncertainty range. 

{To check the reliability of SED parameters and reconstructed histories, we conducted a simulation test with mock data set (Appendix~A). From the test, we found that the scatter in the reproduced amplitude and metallicity at each age bin is $\sim0.5$\,dex and $\sim0.25$\,dex, respectively. To account for this, we add the estimated scatter to the reconstructed star formation and metallicity histories, which are also propagated to other SED parameters (Section~\ref{ssec:hist}).}

\begin{figure*}
\centering
	\input{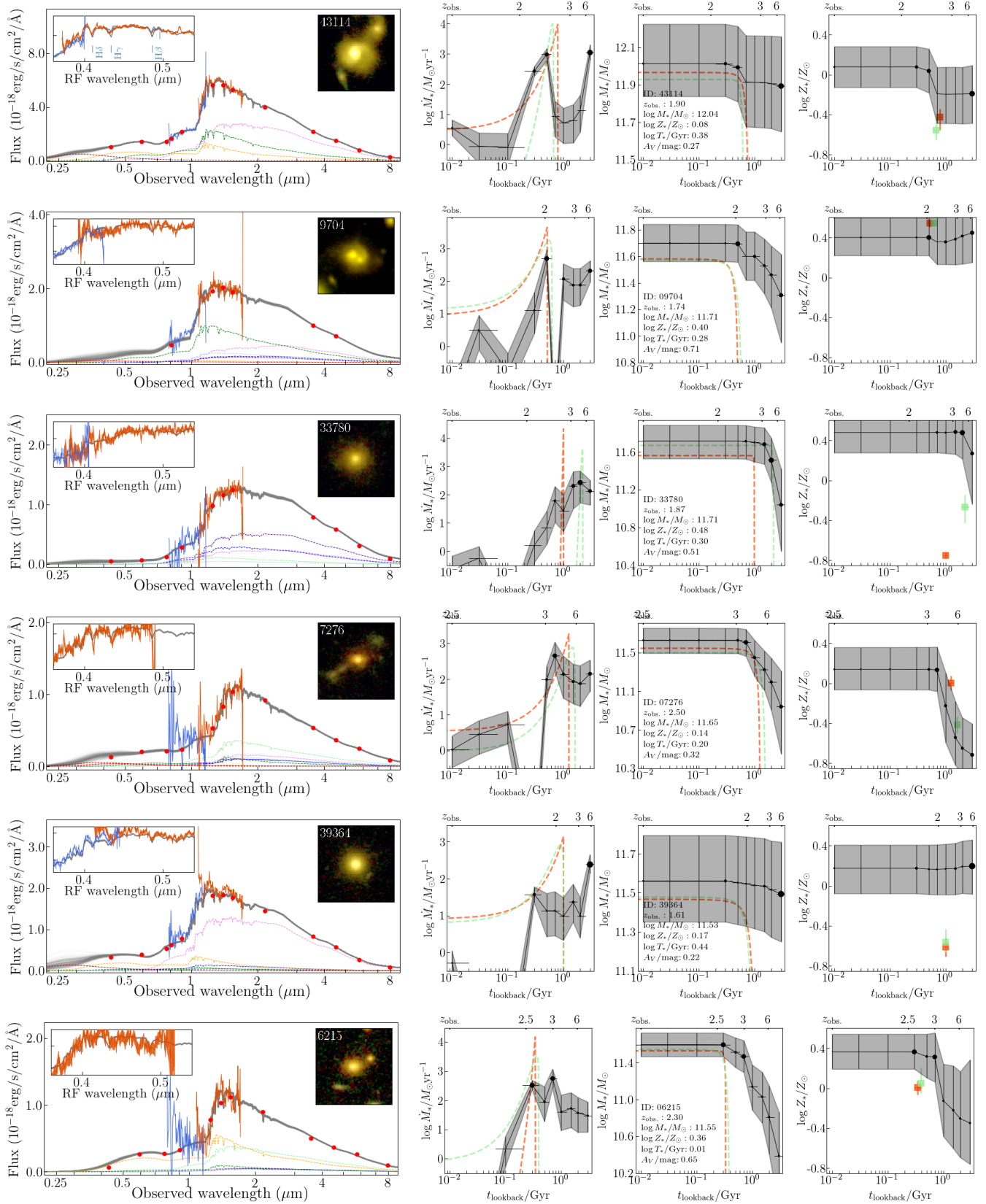}
	\caption{
	Results of SED fitting for the final \Ns\ galaxies sorted in order of stellar mass.
	{\bf First column:} observed spectra (blue and red lines with error bars for G102 and G141) and photometric data points (red circles with black error bars). The best-fit templates of each age bin (dashed lines, colored as in Figure~\ref{fig:demo}) and the sum (gray dashed lines) are shown. Different total templates randomly reproduced from the MCMC chain are also shown. Pseudo-color stamps (F160/125/814W for R/G/B; {$4.8\times4.8$\,arcsec$^2$}) are shown. Emission lines, when detected, are shown with fitted Gaussian curves in the inset (blues solid lines).
	{\bf Second column:} reconstructed star formation histories, with the 50th (black circles) and 16th/84th percentile ranges (gray hatched region) shown. The size of the symbols at each age bin (black circles) represents the amplitude. The best-fit SFHs with functional forms ($\tau$ and delayed-$\tau$ models) are also shown for comparison (red and green dashed lines).
	{\bf Third column:} same as second column, but for stellar-mass accumulation histories. 
	{\bf Fourth column:} same as second column, but for mass-weighted stellar metallicity. The best-fit metallicities derived by the two functional models are shown at the best-fit age (red/green squares with error bars).
	}
\label{fig:all}
\end{figure*}

\begin{figure*}
\figurenum{3}
\centering
	\input{SFH_plot2}
	\caption{Continued.}
\end{figure*}

\begin{figure*}
\figurenum{3}
\centering
	\input{SFH_plot3}
	\caption{Continued.}
\end{figure*}

\begin{figure*}
\figurenum{3}
\centering
	\input{SFH_plot4}
	\caption{Continued.}
\end{figure*}

\section{Results}
\label{sec:result}

\subsection{Twenty-Four Galaxies as the Final Sample}\label{ssec:sed}
In Figure~\ref{fig:mz}, we summarize basic parameters of final sample galaxies. The final sample consists of \nmacsf, \ngdnf, and \ngdsf\ galaxies from M1149, GDN, and GDS, after the rest of initial samples are visually discarded because of poor redshift fitting quality (Section~\ref{ssec:mcmc}). Due to the partial coverage of the grism observations, and also random contamination from neighboring galaxies, our samples have nonuniform exposure time in the grism observation and signal to noise ratio (S/N; Table~\ref{tab:sed}). The two galaxies in M1149 are those previously reported by \citet{morishita18}. 

In panels (a) and (b) of Figure~\ref{fig:mz}, we plot the distribution of our sample galaxies in redshift--stellar mass/F140W--magnitude spaces, respectively. 
Due to {the increasing sensitivity of G141 grism with wavelength, galaxies at higher redshift are fainter in F140W than those at lower redshift.}
In (c), we show the UVJ color-color diagram for diagnosing galaxy quiescence \citep[e.g.,][]{williams09}. While most of the sample locates within the passive category, there are six galaxies that fall below the passive/star forming boundary in $U-V$. These galaxies are in transition between star forming ($U-V\simlt1$) and passive ($\simgt1.4$), akin to the green valley (or ``quenching") galaxies in the local universe \citep{kauffmann03, schawinski14}.

The characteristic age, which is about the time since a system reached its half-mass, of each galaxy represents the mass-weighted value from the reconstructed SFH,
\begin{equation}\label{eq:tm}
T_{*} = \sum_{i} t_i  a_i  \Psi_{i} / \sum_{i} a_i  \Psi_{i} 
\end{equation}
where $t_i$ is the median age, $a_i$ best-fit amplitude, and $ \Psi_{i}$ mass-to-light ratio of $i$-th template. An estimated error for each amplitude from the simulation test (Appendix~A) is added in quadrature. The typical error in mass-weighted age is $\sim0.2$\,dex.

The derived parameters are summarized in Table~\ref{tab:sed}. It is noted that estimated errors in stellar mass are larger ($\sim0.2$\,dex) than those listed in original catalogs \citep[$\simlt0.1$\,dex;][]{skelton14}. This is due to the fact that our fitting accounts for an additional uncertainty originated in the flexibility of SFHs. While one can implement this by e.g., repeating a SED fitting analysis with different SFHs \citep[][]{wuyts12,morishita15}, the stellar mass measurement in the original catalog (as well as many others) is based on one functional form for SFHs, and thus the quoted errors are solely from photometric error and redshift (see also Appendix B).

\subsection{Diagnostics from the SED shape}\label{ssec:spectra}
In the left panels of Figure~\ref{fig:all}, we show SEDs of the sample galaxies with the best-fit templates. Our galaxies are well characterized with $k+a$ \citep{dressler83} and quiescent spectra, as is expected from the sample selection and redshift range. Deep spectra successfully capture spectral features of these types of galaxies, such as absorption lines and $4000\,{\rm \AA}$/Balmer break. The wide broadband coverage well captures spectral features, such as a blue UV slope from a young population ($\sim1$\,Gyr) and near-IR excess from an old population ($\simgt2$\,Gyr), that is consistent with the derived mass-weighted age ($T_*$; see Table~\ref{tab:sed}).

Six galaxies have moderately detected ($\sim1.5\sigma$) weak emission lines, such as \oii, \hd, \hg, and \hb, which is a signature of ongoing star formation. We first fit each emission line with a Gaussian after subtracting the best-fit spectrum. The EW is then measured with the total flux from the Gaussian fit and the best-fit template as a continuum. For \oii\ line, we use Eq.3 in \citet[][]{kennicutt98} to estimate the star formation rates. For \hb\ line, we use assume a recombination coefficient in the Case B \citep{osterbrock89} and then Eq.2 in \citet{kennicutt98}. These weak emission lines indicate specific star formation rate (SFR/$M_*$) of $\simlt 10^{-10}$\,yr$^{-1}$. While the detection is tentative, the low level of star formation activity is also observed in previous findings \citep[e.g.,][]{belli17}, possibly providing more detailed pictures of quenching mechanisms (Section~\ref{sec:disc}). Two of the emission detected galaxies (IDs 19341 and 19850) have strong \oiii\ lines ($4959+5007\,{\rm \AA}$) with a relatively weak \hb\ line, suggesting existence of active galactic nuclei (AGNs). While $\ha$ and $\nii$ are beyond our wavelength coverage, the line ratio of $\hb$ and $\oiii$ lines ($\log \hb/ \oiii\ = 0.15\pm0.03$ and $0.13\pm0.05$) implies that these galaxies as AGNs in the mass-excitation diagram \citep[][]{juneau14}.

On the other hand, we find six galaxies that consist of very old populations with mass-weighted ages $\simgt2$\,Gyr, {and dominate high-mass end among our sample.} While such massive galaxies are rare \citep[$n\sim3\times10^{-5}$\,Mpc$^{-3}$;][]{muzzin13, tomczak14}, it is also true that some of ancient galaxies, that formed a long time ago, have more chance to experience ex-situ processes, e.g., merger and gas accretion, especially at this high redshift. Such ancient galaxies would be smuggled into younger population and become indistinguishable when seen with e.g., light-weighted age. Our reconstructed SFHs have ability to investigate this.

\subsection{SFHs}\label{ssec:hist}
In the right three columns of Figure~\ref{fig:all}, we show the reconstructed SFHs, mass accumulation histories, and metallicity enrichment histories for our sample galaxies. We reconstruct SFRs in each time bin by dividing the amount of stellar mass formed (including the lost mass by the time of observation) by the bin length. Thus, SFRs at each bin represent its average values over the time ($\sim30$\,Myr for the youngest template to $\sim1$\,Gyr for the oldest one). It is also noted that the derived SFR cannot distinguish between to in-situ (stars formed in the system) or ex situ (those obtained via mergers).

Some galaxies are worth highlighting. For example, ID~43114, the most massive galaxy among our sample ($\logm\sim12$; also \citealt{vandokkum10} and \citealt{ferreras12}), formed about 50\% of the final mass already at $z\simgt5$ ($\sim2$\,Gyr ago). The galaxy was at low star formation activity for $\sim1.5$\,Gyr, and then started active star formation ($\sim1000\,M_\odot$/yr) at $z\sim2.5$, $\sim300$\,Myr ago. The significant star formation is comparably high as those of sub-millimeter galaxies at this redshift \citep{younger07,tacconi08}. In fact, its morphology shows a tidal feature with two close objects at the outer part, suggesting a recent (major) merger. Its star formation activity seen in the reconstructed SFH is consistent with a typical merger time scale at this redshift \citep[][]{lotz11,snyder17}. Interestingly, the dust attenuation of this galaxy is relatively low ($A_V\sim0.3$\,mag) compared typical sub-millimeter galaxies and starburst galaxies at this redshift \citep[$A_V>3$\,mag;][]{riechers13,toft14}, suggesting post processes might have cleared a large amount of dust.

Other galaxies with clearly disturbed morphology (e.g., IDs 09701 and 00141) also show recent intense star formation at $\simlt1$\,Gyr before $t_{\rm obs}$, that may provide independent constraint in merger time scale and induced star formation activity from follow-up kinematical studies.

Interestingly, many of our galaxies show extended star formation activity to $\sim0.3$\,Gyr before their observed redshifts. This differs from previous understanding of massive early-type galaxies, whose star formation activity was believed to decline rapidly or truncated after forming the bulk of stars in a very short time. While our sample size here is too small to generalize this \citep[see also][]{ferreras19}, and is also possibly biased to a high-surface density population, it is curious to investigate how previously adopted functional form SFHs behave to such extended feature, as well as other features like dual-peak SFHs. 

We repeat the same analysis but with functional forms of SFHs to compare with our SFHs. In Figure~\ref{fig:all}, we show the best-fit SFHs obtained with a functional form for the SFH, $\tau$-model with $\propto A \exp[-(t-t_0)/\tau]$. In addition to $\tau$ and $t_0$, we allocate one parameter for metallicity, and one for dust attenuation. In most cases, it is clear that SFHs derived from \gsf\ cannon be reproduced by the $\tau$-model. For example, the $\tau$-model cannot capture the dual peaks observed in some of our galaxies (e.g., ID43114). The $\tau$-model also fails to capture extended star formation, both at young and old age sides. This is due to the fact that the functional SFH is light-weighted, where the best-fit parameters are more sensitive to differential amounts of light. Such qualitative discrepancy in fact results in quantitative discrepancy in the best-fit parameters, with $\chi^2/\nu$ systematically larger than \gsf\ (see also \citealt{carnall18b}, who argue limitation by a functionally defined SFH). Appendix~B summarize the comparison of the $\tau$-model, and also results with the delayed $\tau$-model.

{Metallicity histories shown in Figure~\ref{fig:all} represent mass-weighted accumulated metallicity,
\begin{equation}
Z_{j} = \sum_{i}^{i\leq j} Z_{i} a_i \Psi_{i} / \sum_{i}^{i\leq j} a_i  \Psi_{i},
\end{equation}
where $i$ covers older age pixels than $j$. Individual metallicity values at each age pixel often suffer from large uncertainty ($\sim0.25$\,dex; Appendix A), while this is partially attributed to small amount of light in those age pixels (i.e. small $a_i$). We therefore avoid discussing metallicity enrichment histories of {\it individual} galaxies, but rather focus on (more robust) total metallicity in this study (Section~\ref{ssec:Zh}). It is noted, however, that having parameters for metallicity at each age bin allows flexibility in fitting, and more reasonable estimate (i.e. larger error bars) in SFHs and SED parameters.}

\begin{figure*}
\centering
	\includegraphics[width=0.95\textwidth]{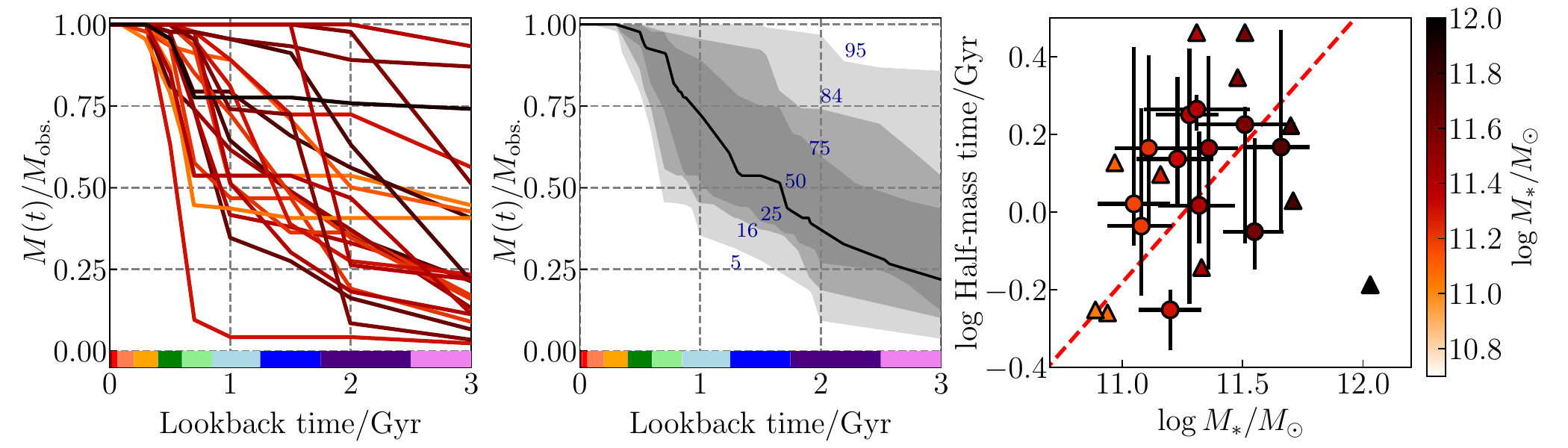}
	\caption{
	SFHs as a function of lookback time from the observed redshift. 
	{\bf Left:} cumulative stellar mass evolution of individual galaxies. Color corresponds to the final stellar mass. Ages that correspond to the SED templates are indicated with color bars at the bottom (colored as in Figure~\ref{fig:demo}). 
	{\bf Middle:} summary of individual cumulative stellar mass evolution, where contour boundaries and line correspond to the 5/16/25/75/84/95th percentiles and median.
	{\bf Right:} half-mass time (lookback time from $t_{\rm obs}$; $t_{50}$ in the main text) as a function of observed stellar mass. Those with lower limits are shown with triangles. Symbol color corresponds to the final stellar mass. A positive correlation, $\log t_{50}$\,/Gyr\,$\propto 0.5 \logm$, is seen (red dashed line; i.e. downsizing).
	}
\label{fig:sft}
\end{figure*}

\section{Discussion}
\label{sec:disc}

\subsection{Time scale of star formation}\label{ssec:sfhsum}
In Figure~\ref{fig:sft}, we summarize the mass accumulation histories as a function of lookback time from the observed redshift. Most of our sample galaxies formed $>50\%$ of their extant mass by $\sim1.5$\,Gyr prior to the observation ($z\sim 2.5$ to 5, depending on the observed redshift), which is quantitatively consistent with recent studies at similar and higher redshift \citep{dominguez16,schreiber18,carpenter18,belli19}.

We estimate the half-mass time, $t_{50}$ (lookback time from the observed redshift), from individual SFHs in the right panel of Figure~\ref{fig:sft}. $t_{50}$ is estimated in each step of MCMC, and thus its uncertainty represents those in SFHs and also individual age bin widths. For some galaxies, we only estimate the lower limit, as $>50\%$ of stellar mass is in the oldest template. Higher resolution spectra by e.g., JWST are required to reveal ancient histories at a higher time resolution. 

Still, we see a trend where more massive galaxies form earlier, known as downsizing \citep{cowie96,heavens04,treu05}, with a linear fit of $\log t_{50}$\,/\,Gyr\,$\propto 0.5 \logm$. The measured standard deviation ($\sim0.16$\,Gyr) is comparable to the redshift range of our galaxies ($\Delta \log T \sim0.14$\,Gyr). The fact that the downsizing trend exists in the early time of the universe provides hints to the galaxy evolution at even earlier epochs, when they were star forming galaxies, and how observed luminous galaxies form \citep{zitrin15,oesch16} in relation to, e.g., their environments \citep{harikane19}.

\begin{figure}
\centering
	\includegraphics[width=0.49\textwidth]{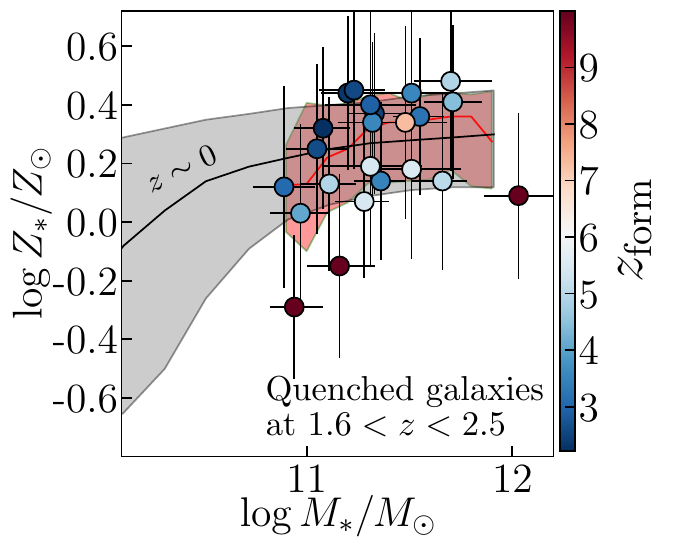}
	\caption{
	Stellar mass-metallicity relation of \Ns\ galaxies in this study (circles). 
	The symbols are color-coded by formation redshift, $z_{\rm form}$. Running median ($\sim0.25$) is shown with 16-84th percentile range (red line with hatched region; $\sigma_{\log Z}\sim0.15$). Most of our galaxies at $z\sim2$ already have consistent values to the local metallicity value (gray hatched region represents 16-84th percentile range from \citealt{gallazzi05}, with a calibration shift for $+0.15$\,dex; Section~\ref{ssec:MZR}). 
	}
\label{fig:zh}
\end{figure}

\subsection{Stellar Mass--Metallicity Relation at $z\sim2$}\label{ssec:MZR}
The stellar mass-metallicity relation is a key diagnostic of galaxies' chemical and mass maturation histories. The relation encodes the coevolution of stellar mass and chemical enrichment among galaxies, and provides an independent clue to the past evolution than SFHs. The relation is known to hold from the local universe \citep{gallazzi05, panter08, delgado14}, in a wide range of mass \citep{kirby13}, up to $z\sim1$ \citep{gallazzi14,choi14,leethochawalit18}. Beyond the redshift, however, it is still studied with a small sample of galaxies and not clear \citep{onodera12,onodera15,kriek16,morishita18}. While the relation in gas-phase metallicity at $z\simgt1$ may suggest the relation may remain universal to higher redshift \citep{tremonti04, mannucci10, kewley08, maiolino08, yabe14, zahid14, onodera16, wang17}, the observed scatter is still large, due to both a selection bias, different tracer of metallicity, and different physical state of gas phase metallicity \citep[e.g.,][]{andrews13}. We here overview the relation at $z>1.6$ for the first time.

In Figure~\ref{fig:zh}, we show the distribution of our galaxies in the stellar mass-metallicity plane. The metallicity here is a mass-weighted value as for the age (Equation~\ref{eq:tm}). While no significant mass dependency of the metallicity is observed, this is because our galaxies occupy the high-mass end and do not span a wide mass range. In fact, the flattening behavior at a high-mass range is consistent with the relation at lower redshift \citep{gallazzi05,gallazzi14}. The observed metallicity is significantly high (median of $\logZ\sim0.25$) and tight (scatter of $\sim0.15$-0.25\,dex around the median), which is comparable to the observed scatter in the local relation.

While challenging, it is still of particular interest to compare our metallicity measurement with the local relation. To do this, we need to calibrate the absolute value of metallicity. Although \citet{gallazzi05}'s stellar-phase metallicity measurement is based on the total metallicity as in this study (as opposed to an element abundance-based measurement), there is a systematic difference due to the adopted isochrones (MIST versus Padova), each of which has a different definition of solar metallicity ($Z_\odot=0.0142$ vs. 0.0190). We correct this by applying a $+0.15$\,dex offset to the \citet{gallazzi05}'s measurement.

Another systematics is the $\alpha$-abundance of the template. While the templates used in this studies are set to the solar composition \citep{choi16}, it is not clear, due to the low-resolution of our spectra, if these metallicity is enhanced by $\alpha$-elements, despite its completely different origin from iron \citep[e.g.,][]{thomas05}. In fact, \citet{leethochawalit18} recently found that $\sim0.16$\,dex decrease in iron abundance of massive galaxies at $z\sim0.4$ compared to the local value. Given the time of universe (when iron was relatively deficient) and short time scale of star formation for our galaxies (where $\alpha$-elements are enhanced), the high $\logZ$ values may represent the $\alpha$-enhancement, as is the case for $z\simgt1$ galaxy \citep[][]{onodera15,kriek16}. However, due to the definition of the total metallicity used here and \citet{gallazzi05}, \Z\,$\sim$\,\FH\,+\,0.94 \alp\ \citep{thomas03}, changing the template to the alpha-enhanced ones should result in minor differences in this comparison.

While keeping these systematics in mind, we find that most of our galaxies are already on the local relation, with a median measured for the entire mass range of ${\log Z}\sim0.25$. The scatter around the median is revealed to be small ($\sigma_{\log Z}\sim0.15$). This implies that chemical enrichment of this class of massive galaxies, if not all, has already been completed, within the first $\sim3$\,Gyr of the universe. We revisit this in the following section.

Two galaxies fall below the median relation (IDs 00227 and 24569). The former, which was reported and discussed in \citet{morishita18}, shows rather extended SFHs with small metallicity values over the entire history. According to its undisturbed morphology and gradual mass increase, seen in Figure~\ref{fig:all}, accretion of low-mass satellites (i.e. metal-poor) or late-time star formation triggered by the infall of pristine gas may explain the observed properties, rather than more dramatic episode involving, e.g., major merger. Detailed investigation of its inner structure and chemical composition at a higher angular resolution would provide further insight into its enrichment evolution \citep[e.g.,][]{abramson18b,wang18}.

The other galaxy, on the other hand, shows a rapid assembly of mass. Since the galaxy formed a large fraction ($\sim50\%$) of its current mass at $z\simgt6$, its low metallicity is consistent with the cosmic metal enrichment \citep[e.g.,][]{lehner16} as well as an observed rapid decrease in gas-phase metallicity at a given mass \citep[][]{troncoso14,onodera16}. Given the time left to $z\sim0$, metal-poor galaxies like ID24569 would possibly be enriched in metallicity by, e.g., mergers and recycled gas and may sneak into the local average population \citep[see discussion in][]{morishita18}. 

It is noted that the systematic uncertainty in $\alpha$-enhancement would not explain the small value in $\logZ$, as both iron and $\alpha$-abundances need to be significantly low.

\begin{figure}
\centering
	\includegraphics[width=0.43\textwidth]{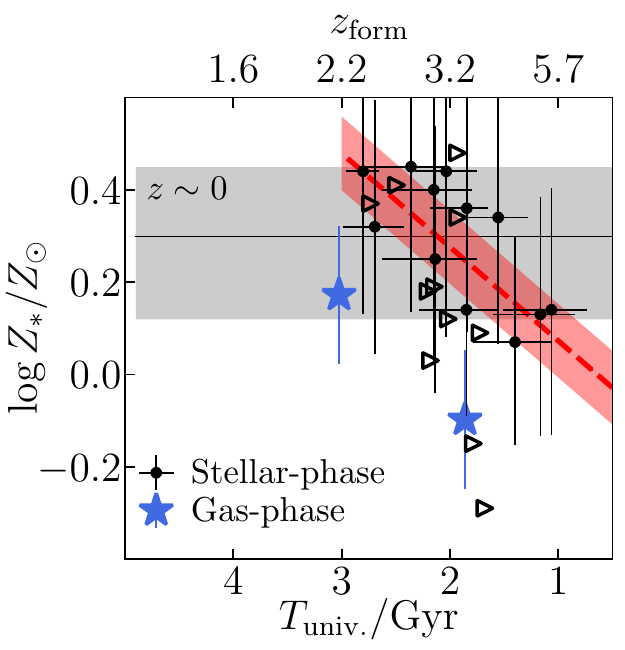}
	\caption{
	Observed stellar metallicity as a function of the formation time, or $z_{\rm form}$ (black circles/triangles for those with an upper limit in age estimate). The linear fit with a slope ($0.20\pm0.08$\,dex/Gyr) and standard deviation ($\sigma \sim 0.16$\,dex) is shown (red dashed line and shaded region). The local metallicity value of massive galaxies \citep[$\logm\sim11.5$;][]{gallazzi05} is shown for comparison (gray shaded region). Gas-phase metallicity measurements of massive galaxies ($\logm=11.5$) at similar redshifts \citep[][]{maiolino08} are shown (blue stars), after being calibrated to the local $Z_*$ measurement (Section~\ref{ssec:Zh}). 
	}
\label{fig:delzh}
\end{figure}

\subsection{Redshift Evolution of Stellar-phase Metallicity}\label{ssec:Zh}
While our mass-metallicity relation indicates that galaxies are already enriched to the value at present day (Figure~\ref{fig:zh}), their origin and observed scatter, especially those with small metallicity, is yet to be investigated. We investigate the redshift evolution of total metallicity as a population by considering the formation time ($z_{\rm form}$), which is derived with the mass-weighted age and observed redshift, $T_{\rm form} = T(z_{\rm form}) = T(z_{\rm obs})-T_*$ (i.e. lookback time to the half-mass time). 

In Figure~\ref{fig:delzh}, we show the distribution of metallicity as a function of $T_{\rm form}$. A clear correlation between $T_{\rm form}$ and observed metallicity is observed. A linear regression reveals a slope of $\log Z_*/dt\sim0.20\pm{0.08}$\,dex/Gyr, with a standard deviation of $\sim0.15$\,dex. Our result shows the metallicity enrichment happening in this class of massive galaxies, whose metallicity already reaches the local value at $z\sim3$, for the first time. 

One may suspect that this is an artifact from the age-metallicity degeneracy. While some galaxies show a weak correlation between the two parameters, the degree of correlation is much smaller than the quoted error bars in Figure~\ref{fig:delzh}. Our simulation test also revealed that the total metallicity/age are reproduced reliably enough for the observe trend (see Appendix~A).

Interestingly, the linear regression suggests that the metallicity even exceeds the local value of the most massive galaxies in \citet{gallazzi05} by $\sim0.1$\,dex. As noted before, our sample galaxies are biased to compact, high density galaxies due to high S/N requirement for the SED fitting. While compact massive galaxies are rare at $z\sim0$ \citep[][but see also \citealt{poggianti13}]{taylor10}, following events such as minor mergers/second bursts in the following $\sim10$\,Gyr would resolve the tension. We discuss this in the following section.

In Figure~\ref{fig:delzh}, we also plot gas-phase metallicity measurements of star forming galaxies in a similar redshift range for comparison. We use the formula derived in \citet{maiolino08} at the same mass ($\logm=11.5$). {We match the gas-phase metallicity measurement to the stellar metallicity at $z\sim0$. While comparing absolute values of two different metallicities is extremely challenging due to a number of uncertainties in each measurement \citep[e.g.,][]{sanchez17,bian18}, the matching process is reasonable for our purpose here, i.e. comparison of relative differences at $z\sim0$ and $2$.}

It is interesting to find an offset of $\sim0.2$-0.3\,dex between those metallicity measurements in our redshift range, that may give us a clue to how massive quenched galaxies enrich metallicity in this redshift. Assuming those star forming galaxies are star forming counterparts of our passive galaxies, the observed gap has to be resolved between half-mass time (i.e. $T_*$ ago) and time when they are observed as quenched galaxies.

One possible explanation is the continuation of a low-level star formation activity in a closed box, or ``strangulation" \citep{larson80,peng15}. \citet{peng15} demonstrated in their chemical evolution model that the observed offset in metallicity of local massive star forming and passive galaxies ($\sim0.1$\,dex) can be explained by strangulation. Rapid cessation of star formation by AGN/stellar feedback would instead reproduce a similar metallicity for the two populations. Recent observations in fact revealed low star formation activity \citep[e.g.,][]{gobat17,belli17}, as well as remaining gas \citep[][but also \citealt{sargent15,bezanson19}]{gobat18}, in massive quiescent galaxies at $z\sim2$, implying the continuation of star formation after they have formed a large amount of stars, or quenched. In addition, the closed-box enrichment seems to be a good agreement with the observed individual SFHs (Figure~\ref{fig:all}), independently supporting our speculation here.

However, more detailed chemical modelings would be required to reach a conclusion. For example, the observed gap may also be attributed to the dilution of gas-phase metallicity by infalling pristine gas, while there is no gas infall in quenched galaxies due to virial shock heating \citep[e.g.,][]{birnboim03}. If this is the case, it is suggested that galaxy quenching may be largely caused by termination of gas infall \citep[e.g.,][]{feldman15}, while it is not likely that cutting the gas supply would result in extended SFHs as we observe here. A sophisticated chemical modeling with a panchromatic data set, including gas-mass measurements, would be required for further understanding.

\subsection{Following Evolution to $z\sim0$}\label{ssec:follow}
We have found that our galaxies are already enriched in metallicity, located on the local mass-metallicity relation. Given the amount of mass and its quiescence, it is within our interests to investigate how these galaxies will evolve to the local population. While there is a large uncertainty to expect their descendant (as described in the Introduction), it is still worth describing their possible paths and mechanisms.

In particular, many members of our sample are compact, high-density galaxies ($\langle r_{\rm eff}\rangle \sim2$\,kpc; possibly due to the selection bias toward high S/Ns). Compact galaxies at these redshifts are often debated in terms of size evolution, where observed size is $\sim3$--5 times smaller than galaxies at $z\sim0$ with similar masses \citep{trujillo07, vanderwel14, morishita14}. While there is still much debate as to whether (all of) these galaxies would follow such a significant size evolution \citep{nipoti12, newman12, poggianti13,belli17}, minor merger is a popular mechanism that can efficiently increase their sizes \citep[e.g.,][]{naab09, hopkins09, oser10, vandokkum15b, morishita16}. 

The scenario appears to consistently work for our result of metallicity, where accretion of low-mass galaxies (which are less-metal-enriched, expected from the mass metallicity relation) would dilute the system's metallicity to the consistent value. For example, approximately 5 minor mergers, with 1/10 the mass of the host and metallicity inferred from the relation at $z\sim2$, would lower the host metallicity by $\sim0.1$\,dex, being consistent with the local value. The metallicity gradient observed at $z\sim0$ \citep[e.g.,][]{delgado14,martin-navarro18} is independent evidence that such high-$z$ metal-rich galaxies would become cores while the accreted component locates the outer part of local massive galaxies. The integrated metallicity is instead an average value of the whole system; thus, metallicities observed in the local relation should be lower than those observed at higher redshifts.

Infall of metal poor gas associated with minor merging satellites \citep[e.g.,][]{torrey12} or direct infall from the cosmic web \citep{dekel06} would also dilute the system's total metallicity by inducing the second burst. While it is not clear if the scenario reproduces the observed metallicity gradient at $z\sim0$, there is a large fraction of early-type galaxies that show an evidence of ongoing star formation at the intermediate redshift \citep[e.g.,][]{treu05b, kaviraj11}. Spatially resolved studies of such second burst galaxies will shed light on how these different scenarios contribute to the evolutionary path of massive galaxies at high redshift to the local counterpart.

\section{Summary}
\label{sec:sum}
We reconstructed SFHs of \Ns\ massive, passively evolving galaxies at $z\sim2$. Our new SED modeling with \gsf\ simultaneously fit slitless spectroscopic and photometric data taken from multiple surveys, with no functional assumption for SFHs. Our main findings are as follows.

\begin{enumerate}
	\item Our massive galaxies have already formed $>50\%$ of their current mass by $\sim1.5$\,Gyr prior to the epoch of observation, with a downsizing trend where more massive galaxies evolve earlier.
	\item The SFHs reconstructed by \gsf\ show a more extended feature than what is obtained with a $\tau$-model fitting for most of the sample galaxies, indicating a low-level star formation activity until recently, rather a than abrupt cessation.
	\item The stellar-phase metallicities of most of our galaxies are already compatible with local values, indicating a rapid metallicity enrichment being associated with the early stellar-mass formation.
	\item By using the reconstructed SFHs and inferred metallicity, we revealed a rapid metallicity enrichment of this class of massive galaxies at a rate of $\sim0.2$\,dex/Gyr in $\logZ$ from $z\sim5.5$ to $2.2$.
	\item While systematic uncertainties remain, the observed gap between the stellar- and gas-phase metallicities can be explained by continuation of a low-level of star formation in quiescent galaxies and/or dilution of gas-phase metallicity due to the inflow of pristine gas to star forming galaxies. The former scenario is consistent with the finding from individual SFHs.
\end{enumerate}

\acknowledgements
We thank the anonymous referee for reading the manuscript carefully and providing constructive comments. We thank Marco Chiaberge, Colin Norman, Kartheik Iyer, Sara Ellison, and Susan Kassin for fruitful discussion. We thank Benedikt Diemer for providing SFHs extracted from the Illustris simulation. We thank Pascal Oesch for providing their HDUV data prior to the public release. Support for GLASS (HST-GO-13459) was provided by NASA through a grant from the Space Telescope Science Institute, which is operated by the Association of Universities for Research in Astronomy, Inc., under NASA contract NAS 5-26555. Support for this work is provided by NASA through a Spitzer award issued by JPL/Caltech, HST-AR-13235 and HST-GO-13177. MT acknowledges the support provided by the Australian Research Council Centre of Excellence for All Sky Astrophysics in 3 Dimensions (ASTRO 3D), through project No. CE170100013. 

{\it Software:} Astropy \citep{muna16},  {\it emcee} \citep{foreman13}, lmfit \citep{newville17}, Grizli \citep{brammer18}, python-fsps \citep{conroy09fsps,conroy10, foreman14}

\begin{figure*}
\centering
	\includegraphics[width=0.9\textwidth]{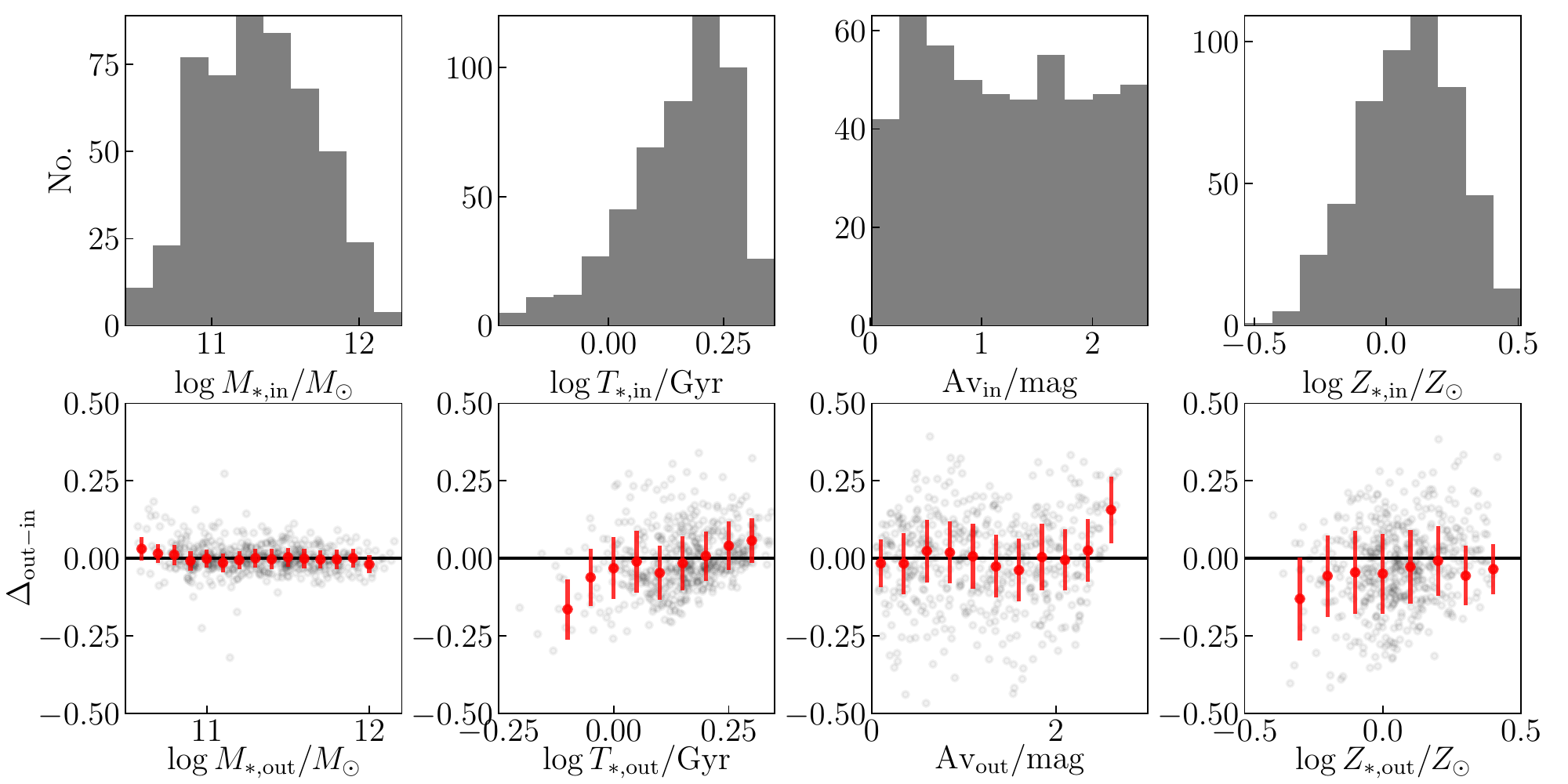}
	\caption{
	{\bf Top:} Parameter distributions of 600\,mock galaxies. Noted that age is distributed in a linear scale (peaked at $\sim1.5$\,Gyr), but shown in a log scale here.
	{\bf Bottom:} offsets of input and output values ($\Delta_{\rm out-in}$; solid lines represent zero point). Median offsets and 16/84th percentiles range at each output bin are shown (red circles and error bars). Stellar mass, dust attenuation, and mass-weighted metallicity show good agreement for the parameter ranges of our galaxies. While a weak positive correlation is seen in mass-weighted age, the measurement is not biased and scatters are small ($\sigma\sim0.1$) at the median value of our sample ($\log T_*/\,$Gyr$\sim0.2$).}
\label{fig:sim}
\end{figure*}

\begin{figure}
\centering
	\includegraphics[width=0.45\textwidth]{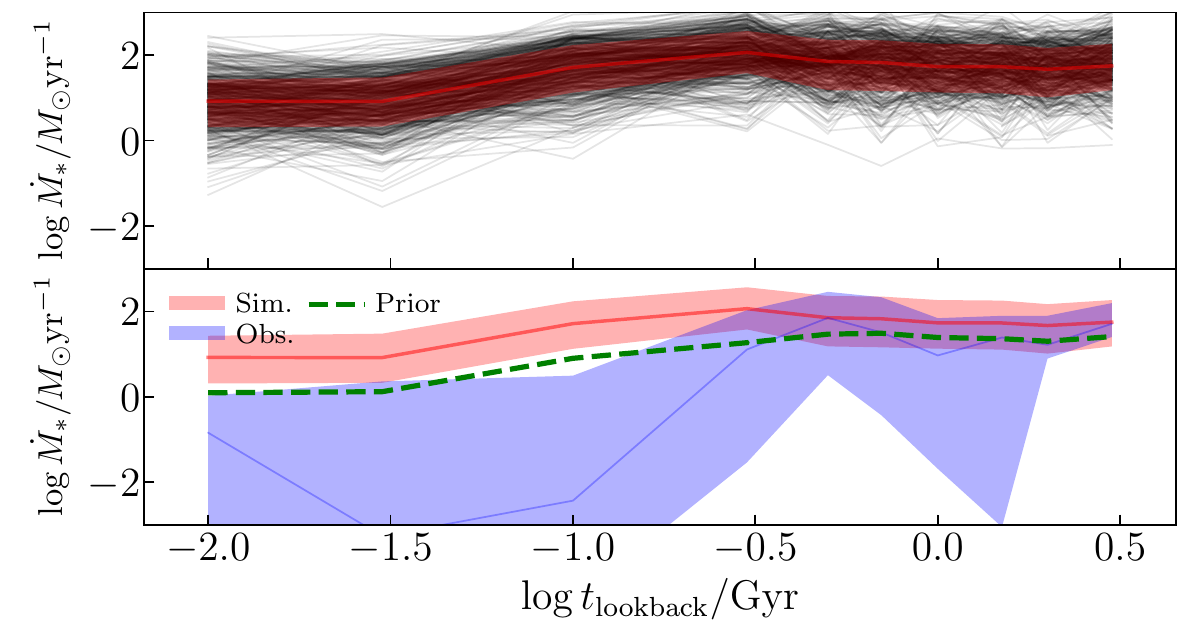}
	\caption{
	({\bf Top}) Input SFHs (thin lines) used for a simulation in Appendix A, with 16/50/84th percentiles (red hatched region).
	({\bf Bottom}) Median input SFH is compared with the median of observed SFH (blue). Since the simulated SFHs are randomly generated, it does not follow the observed decline at $<0.3$\,Gyr.
	The flat prior used in \gsf\ is shown (dashed line). Its effect should be minimal due to the wide constrain range set in this study ($\pm3$\,dex), as seen in the posterior SFHs deviating from the prior.
	}
\label{fig:sfh_rand}
\end{figure}

\begin{figure}
\centering
	\includegraphics[width=0.48\textwidth]{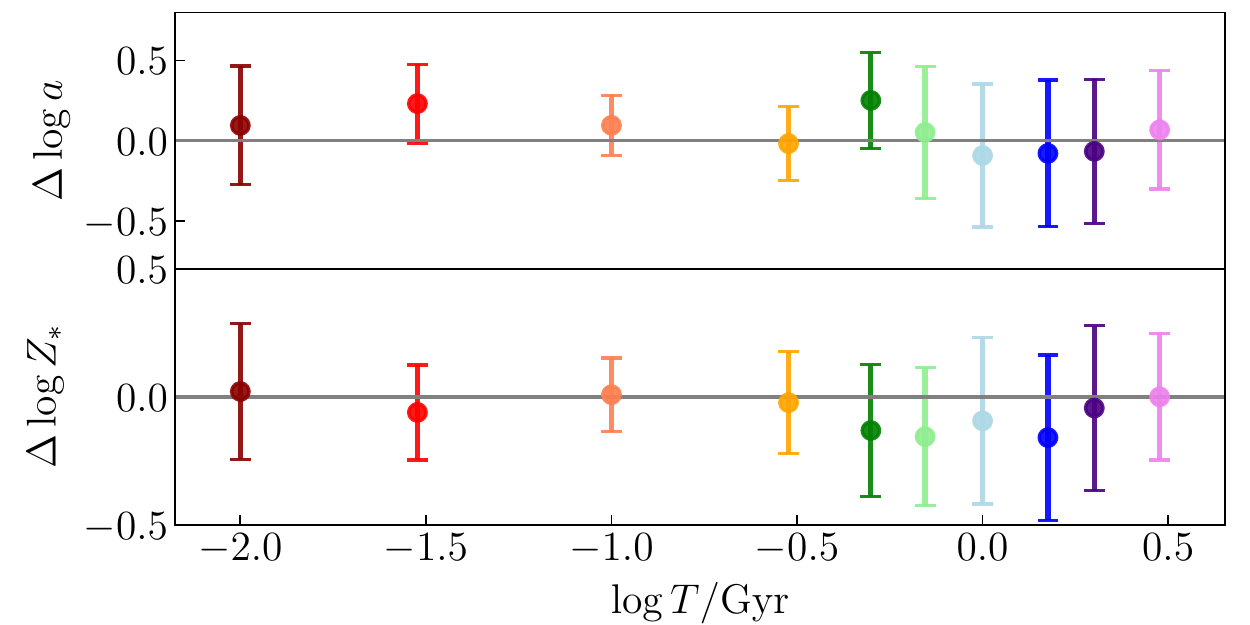}
	\caption{Mean offsets for amplitude ($\log a_{i,{\rm out}}-\log a_{i,{\rm in}}$; top panels) and metallicity ($\log Z_{i,{\rm out}}-\log Z_{i,{\rm in}}$; bottom panels) with standard deviation at each age pixel from the mock test. The measured standard deviation is added in quadrature to observed uncertainty (i.e. star formation and metallicity histories).}
\label{fig:simSFH}
\end{figure}

\section*{Appendix}

\section*{\small Appendix A: Mock Simulation of SED Fitting}\label{sec:Aa}
We test the fidelity of galaxy SFHs and other parameters with our SED fitting method. 

\subsection*{\small A1: Simulation Setup}\label{sec:Aa1}
To explore parameter spaces we are interested in this study (i.e. quenched galaxies at $z\sim2$), we set parameters as follows: redshift $z\in[1.6:2.5]$; mass-weighted age $T_*/$\,Gyr\,$\in[0.6:2.2]$ peaked at $\sim1.5$\,Gyr; dust attenuation $A_V/$mag$\in[0:2.5]$ with a flat distribution; and metallicity $\logZ \in[-0.5:0.5]$, peaked at $\logZ=0.15$\,dex (top panels of Figure~\ref{fig:sim}). The amplitudes of each template ($a_i$; i.e. SFHs) are randomly assigned. The input SFHs are shown in Figure~\ref{fig:sfh_rand}.

The mock SEDs are generated via FSPS \citep{conroy09fsps,conroy10} with the assigned parameters. While we provide SFHs at the same time resolution as the fitting templates, this turns out only a small effect, thanks to a sufficient number of age bin (see also Appendix C for the result with higher-resolution SFHs).

Broadband photometry is then extracted by convolving the mock SEDs with filter response curves. For grism spectra, we convolve the mock template with observed line spread function (modeled with a gaussian), which takes into account morphology/instrumental convolution. The error of each spectral element and broadband photometry is randomly assigned based on the observed uncertainty of 24 galaxies in this study ($\sim5$-20/\,pixel at $3700<\lambda_{\rm rest}/{\rm \AA}<4200$; Table~\ref{tab:sed}). 

No emission lines are added since our focus is the quenched/old galaxy population. In total, 600 mock sets of G102/G141 spectra+broadband photometry are prepared from the template generated with random sets of parameters. We follow the same fitting method as in the main text.

\begin{figure*}
\centering
	\includegraphics[width=0.37\textwidth]{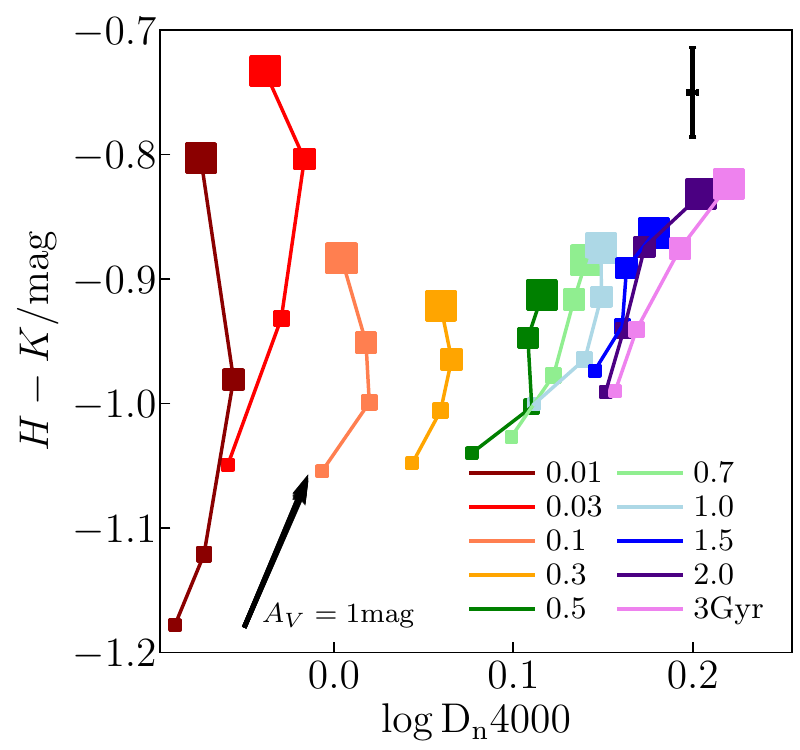}
	\includegraphics[width=0.6\textwidth]{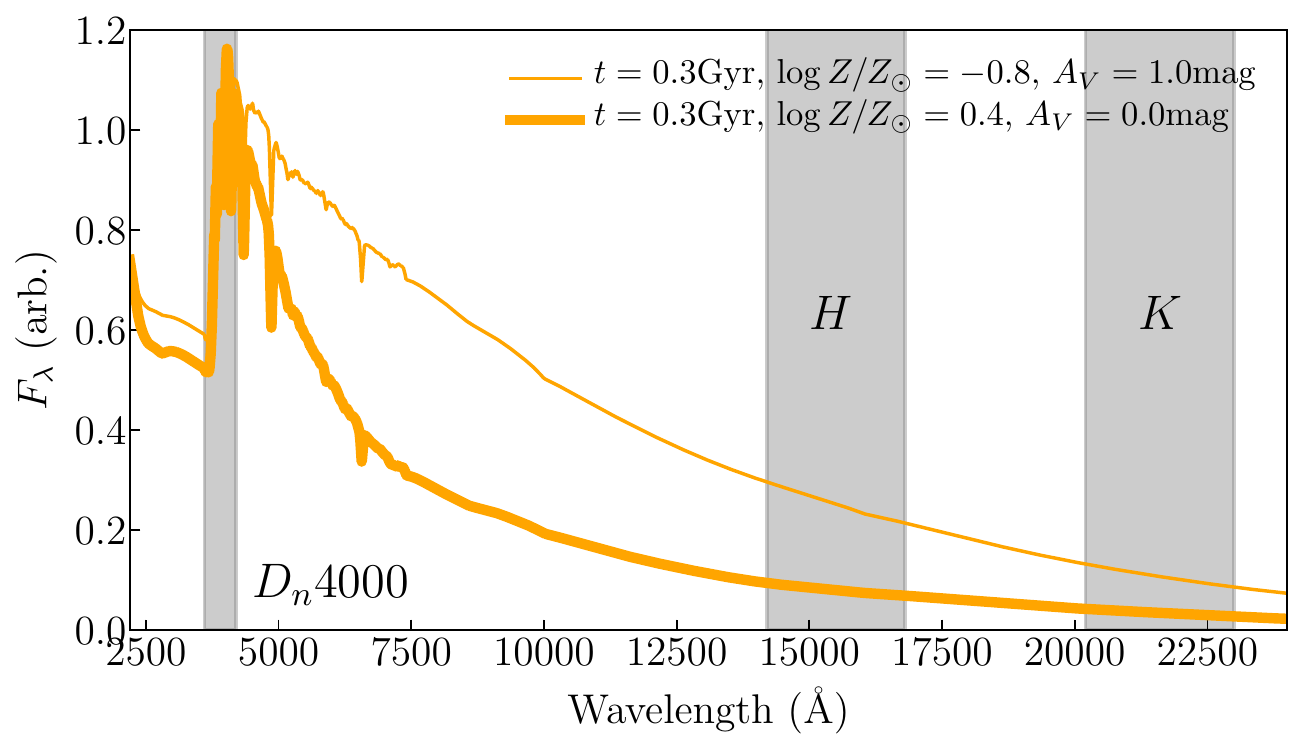}
	\caption{
	(Left) Rest-frame NIR color ($H-K$) as a function of \Dn, for each age (lines with different colors) and metallicity ($\logZ=-0.8, -0.4, 0, 0.4$ for small to large symbols). Both parameters are derived from templates that are convolved to a comparable resolution of retrieved grism data ($R\sim150$), though this is a minimal effect here. Age and metallicity are nearly orthogonal in most of the parameter space, except for the old and solar/supersolar metallicity population at the top right. Typical uncertainties in $H-K$ color and \Dn\ are shown at top right.
	(Right) example of two spectral templates with similar color and \Dn\ as in the left panel. Despite the similarity, these two templates are distinguishable with photometric data at the optical-to-NIR wavelength range.
	}
\label{fig:colcol}
\end{figure*}

\subsection*{\small A2: Result of Global Parameters}\label{sec:Aa2}
Figure~\ref{fig:sim} shows the offset of output and input values ($\Delta y = y_{out} - y_{in}$) as a function of output value for major parameters---stellar mass, mass-weighted age, dust attenuation, and metallicity of the mock galaxies. By taking output values (rather than input ones) in $x$-axis, it is possible to infer the false-positive fraction at a given output (i.e. observed) value, and also implement the scatter to observed values for more comprehensive estimate of uncertainty.

We find excellent agreement in stellar mass with scatter of $\sim0.05$\,dex, and moderate agreement in mass-weighted age,  dust attenuation, and mass-weighted metallicity with scatters of $\sim0.11$, $0.14$, and $0.13$ around median values. Median offsets are small for most of parameter ranges. Mass-weighted age shows a negative slope, underestimating for $\sim0.15$\,dex at $\log T\simlt-0.05$. However, most of our sample galaxies dominate higher values, with a median of $\log T_*$/\,Gyr\,$\sim0.2$ (Figure~\ref{fig:sft}), where the bias in parameters is small, and thus we do not correct the offset for our galaxies in the main text. 

\subsection*{\small A3: Result of SFHs}\label{sec:Aa3}
In Figure~\ref{fig:simSFH}, we summarize offsets of output and input values for amplitude and metallicity at each age pixel to show the fidelity of SFHs and metallicity enrichment histories from all mock galaxies used here. While the offset and scatter may depend on parameter sets with different combinations, this suggests that SFHs can be determined in $\sim0.5$\,dex accuracy. 

Metallicity shows a large scatter in reproduced values, with standard deviation of $\sim0.3$\,dex. Given the parameter range assigned for metallicity ($\in[-0.8:0.6]$), we conclude that determination of metallicity at {\it each} age pixel is challenging with the current data set. It is noted that this does not mean that reproduced {\it total} metallicity has comparable uncertainty, since part of the scatter can be attributed to the age pixel where the total contribution of light is small. Total metallicity, which is light or mass-weighted, should remain less scattered ($\simlt0.2$\,dex), as shown in the main text and A2.

\begin{figure}
\centering
	\includegraphics[width=0.4\textwidth]{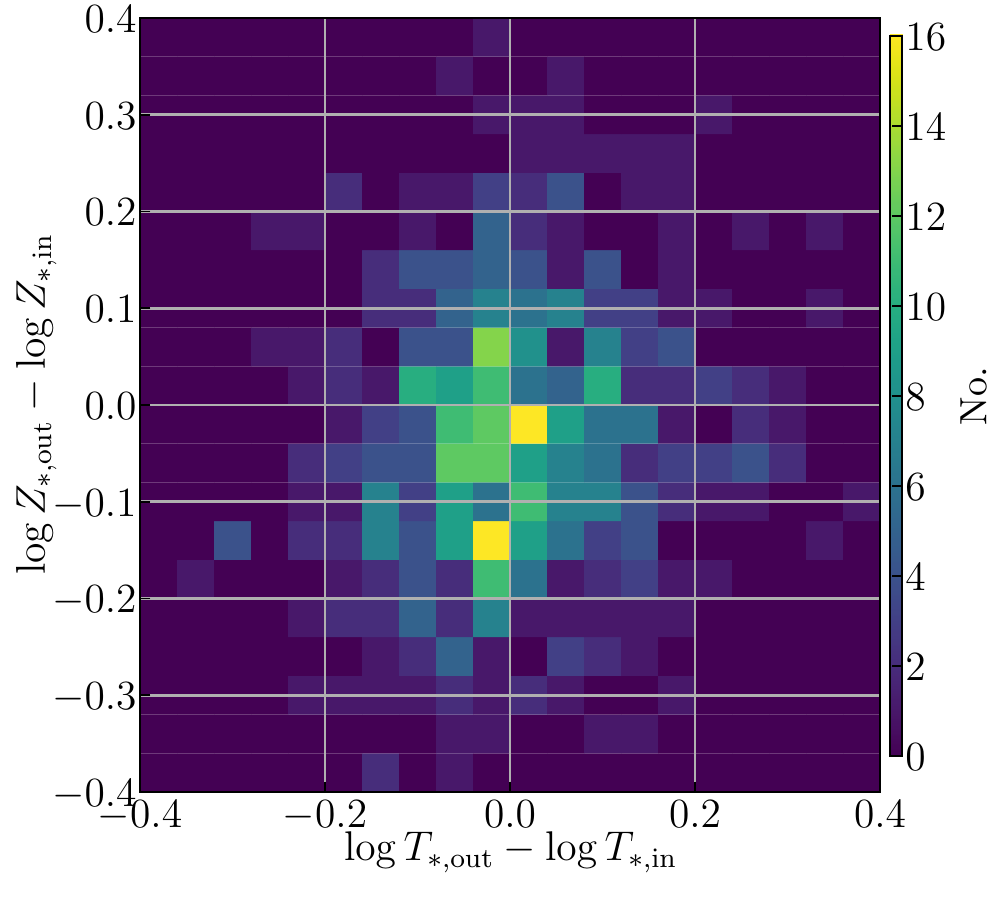}
	\caption{Distribution of offset of age ($\log T_{*,{\rm out}}-\log T_{*,{\rm in}}$; $x$-axis) and metallicity ($\log Z_{*,{\rm out}}-\log Z_{*,{\rm in}}$; $y$-axis) from our mock test. Color represents the number of galaxies in each grid. The distribution is almost symmetric around the zero point, whereas it would follow a negative slope in case of the age-metallicity degeneracy.}
\label{fig:simTZ}
\end{figure}

\subsection*{\small A4: Age-Metallicity Degeneracy}\label{sec:Aa4}
Degeneracy between age and metallicity is notoriously known as one of difficult aspects when modeling accurate SEDs from photometric data \citep{worthey94}. The degeneracy is however resolved once one obtains both information at optical and NIR wavelength range simultaneously \citep[Figure~\ref{fig:demo}; also][]{dejong96,smail01,choi16}.

To see if this is the case for our data set, we first show in the left panel of Figure~\ref{fig:colcol} a rest frame NIR color-\Dn\ diagram, both of which are available with our data in this study. Rest frame NIR color ($H-K$) and the strength of 4000\,\AA\ \citep[\Dn][]{balogh99} are calculated with templates used for fitting, which are convolved to a comparable resolution of grism data ($R\sim150$), including the convolution effect by source morphology. As we see in the figure, age and metallicity are nearly orthogonal in most of the parameter range, meaning the age and metallicity can be well separated from those measurement. The only exception is for old ($\simgt1.5$\,Gyr) and solar/super-solar metallicity, where the relation of the two measurements becomes less orthogonal. Due to this, metallicities of our fitting typically have larger uncertainties for old populations.

{One may notice that some parameters are not distinguished, especially by the reddening effect by dust (e.g., $t=0.3$\,Gyr with $\logZ=0.4$, $A_V=0$ vs. $t=0.3$\,Gyr with $\logZ=-0.8$, $A_V=1.0$\,mag).  However, we stress that to our SED fitting is not relying on any specific colors or indicators, but on all spectrophotometric information over the wide wavelength range. As an example, we show two spectral templates in the right panel of Figure~\ref{fig:colcol}, that locate at a similar position in the \Dn-color space. Despite this, the two templates are clearly distinguishable at rest frame optical to near infrared wavelength range, where sufficient photometric data points are available in this study. Also, photometric error of each flux measurement may affect the SED parameters, but the uncertainty is properly implemented in our fitting framework using MCMC and reflected in the uncertainty range of posterior.}

We also investigate the age-metallicity degeneracy with our mock data set. In Figure~\ref{fig:simTZ}, we show the distribution of offset in total mass-weighted age and metallicity for our mock galaxies. The distribution is symmetric in both axes, whereas the distribution would follow a negative slope if these parameters are degenerated. The distribution is scattered for $\sim0.2$\,dex, which is consistent with those found in A2.

From both tests here, we conclude that the data set used in this study can resolve the age-metallicity degeneracy for our moderately old galaxies ($\simlt2$\,Gyr), but star formation and metallicity histories become less certain beyond $T_{\rm lookback}\simgt2$\,Gyr.

\section*{\small Appendix B: Comparison of SED Parameters Obtained with Functional SFHs}\label{sec:Ab}
In Section~\ref{ssec:hist}, we see that our reconstructed SFHs capture the detail features of individual galaxy SFHs, that are often missed with functional forms. While the deviation is clear in these comparison, it is yet to be investigate how different assumption of SFH results in SED parameters. We here compare the best-fit parameters between two types of SFHs.

\begin{figure*}
\centering
	\includegraphics[width=0.95\textwidth]{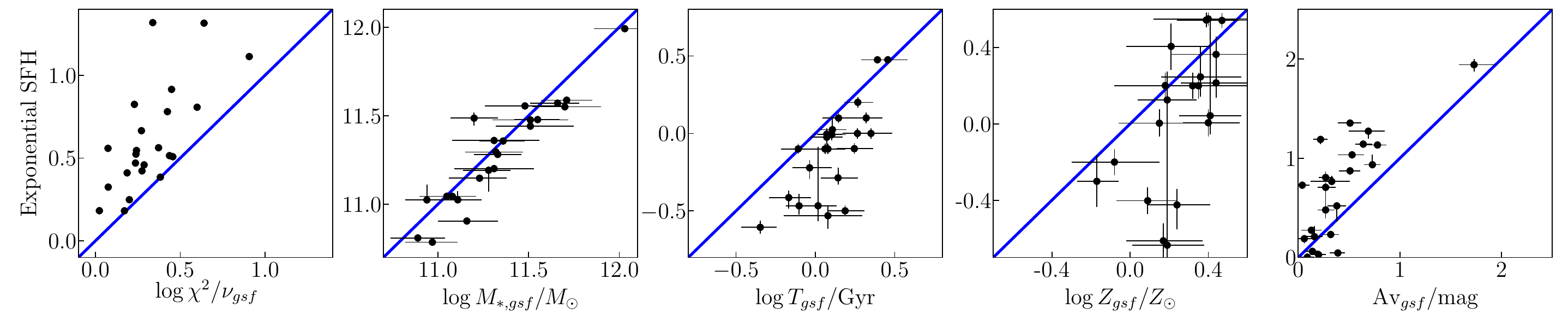}
	\includegraphics[width=0.95\textwidth]{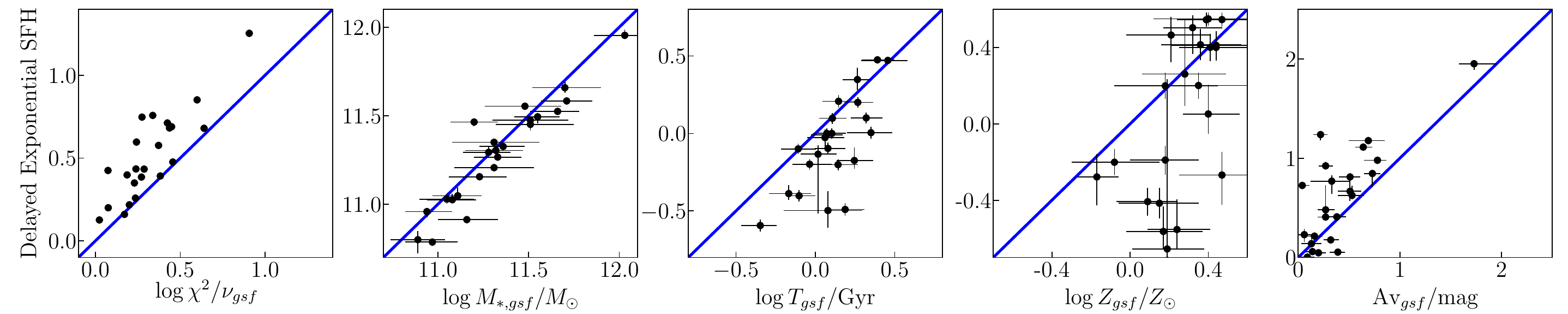}
	\caption{
	Comparison of $\chi^2/\nu$, and SED parameters for \Ns\ galaxies in this study derived from different SFHs, with \gsf\ ($x$-axis) and with functional forms for SFHs ($y$-axis). The exponential model (top) and delayed (bottom) models are examined. The goodness of fit is better with \gsf\ for most of our sample. While both SFHs reproduce the stellar mass in a good agreement, the functional SFHs overestimate dust attenuation and underestimate age and metallicity. It is noted that age and metallicity of \gsf\ are light-weighted values to match those derived with functional form SFHs, while those in the main text are mass-weighted values.
	}
\label{fig:sfhcomp}
\end{figure*}

In Figure~\ref{fig:sfhcomp}, we compare the goodness of fit, $\chi^2/\nu$, and major parameters from SED fitting, i.e. stellar mass, {\it light-weighted} age and metallicity, and dust attenuation. These values are compared between those reproduced by \gsf\ (main text) and functional ones. Firstly, the goodness of fit is better with \gsf\ for most of our galaxies, which is reasonable given the flexibility of its modeling. We also find that the $\tau$-model ($\propto \exp[-t/\tau]$) is more sensitive to the light from young stellar populations, where the model systematically underestimates system's ages for $\sim 0.3$\,Gyr on average (top panels). The discrepancy propagates to other parameters, where we find over estimated dust attenuation ($\sim0.3$\,mag) and largely underestimated metallicity.

The discrepancy in age and dust becomes slightly smaller when the delayed-$\tau$ model ($\propto t \exp[-t/\tau]$) is used (bottom panels of Figure~\ref{fig:sfhcomp}). This is shown in the goodness of fit, where the delayed-$\tau$ model results in smaller values of $\chinu$ \citep[see also][]{pacifici16}. This is partly attributed to its rising slope in SFHs, which makes the age slightly older and cancel out discrepancy in other parameters. A large discrepancy in metallicity, however, still remains.

Despite this, it is interesting that the reproduced stellar mass is very consistent with each other, with only $\sim0.1$\,dex scatter in $\log M_*$ among our sample. While it is challenging to comprehensively understand this agreement because of a large number of parameters here, this is partly due to the sufficient wavelength coverage to the rest-frame near-IR region, where the light from low-mass stars dominates and less sensitive to other parameters \citep[e.g.,][]{bell03}. The contribution from other parameters (i.e. age/metallicity/dust) are cancelled out within partial degeneracy at a given form of star formation history. However, this does not necessarily mean that the typical error in stellar mass remains comparably small. As shown in the main text and Appendix A, the stellar mass measurement with \gsf\ accompanies with $\sim0.2$\,dex uncertainty, that is mainly originated from systematics in estimating accurate star formation and metallicity enrichment histories (cf. smaller uncertainties in stellar mass with functional form SFHs estimated here). For this reason, we conclude that stellar mass measurement remains, at least, $\sim0.15$\,dex accuracy for our galaxies, and perhaps for other types of galaxies. The best-fit parameters derived with the two functional SFHs are summarized in Table~\ref{tab:SFH2}.

\section*{\small Appendix C: Simulation with Realistic SFHs}\label{sec:Ac}
While our test with randomly generated SFHs provides an general idea of the goodness of \gsf, there is still concern of how a specific type of SFHs affects output results. In particular, the random SFHs do not fully investigate SFHs of quenched galaxies, i.e. target galaxies of our studies. Upon such a demand, we here repeat a similar fitting analysis as in Appendix~A, but with SFHs taken from a cosmological simulation, that gives us an idea how well quenched galaxies are reconstructed within our framework.

\begin{figure*}
\centering
	\includegraphics[width=0.49\textwidth]{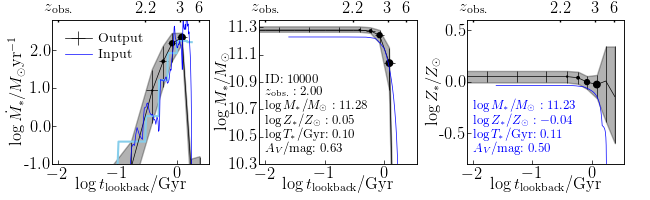}
	\includegraphics[width=0.49\textwidth]{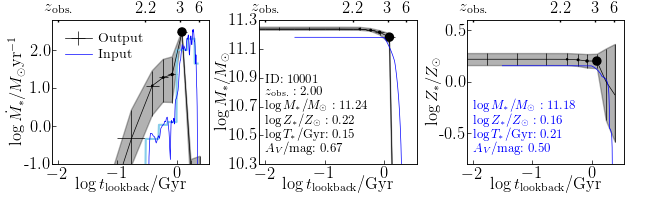}
	\includegraphics[width=0.49\textwidth]{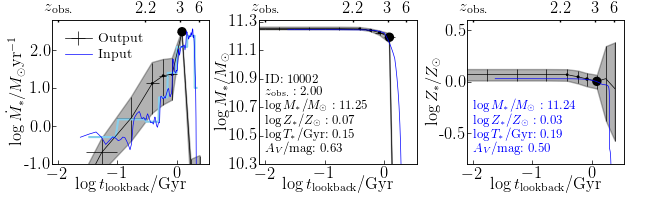}
	\includegraphics[width=0.49\textwidth]{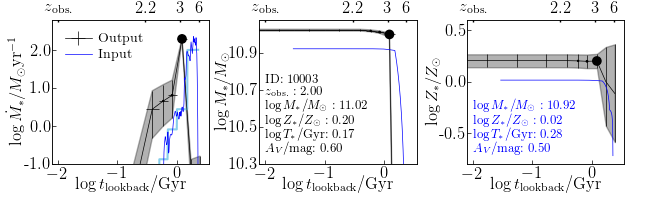}
	\includegraphics[width=0.49\textwidth]{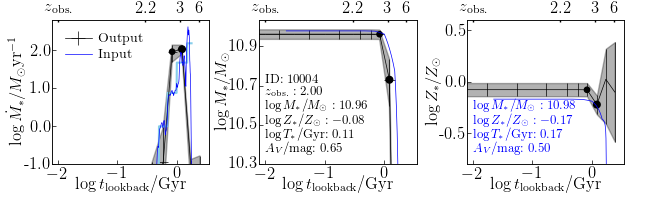}
	\includegraphics[width=0.49\textwidth]{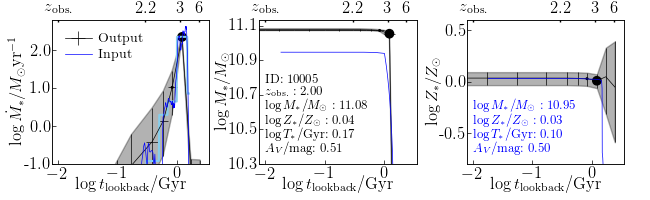}
	\includegraphics[width=0.49\textwidth]{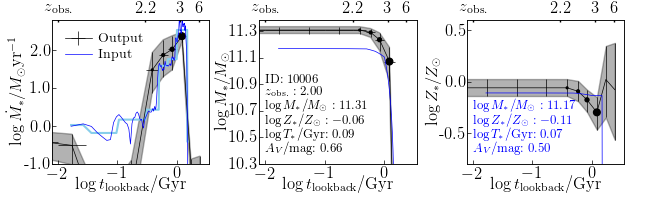}
	\includegraphics[width=0.49\textwidth]{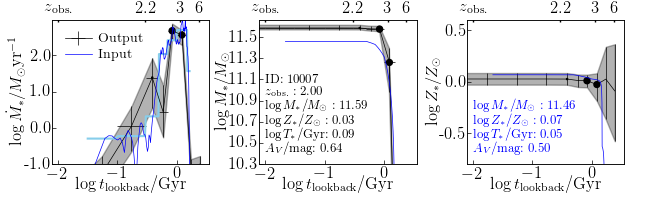}
	\includegraphics[width=0.49\textwidth]{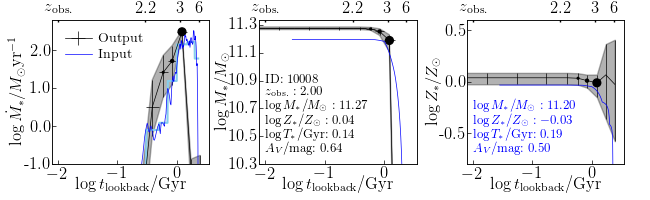}
	\includegraphics[width=0.49\textwidth]{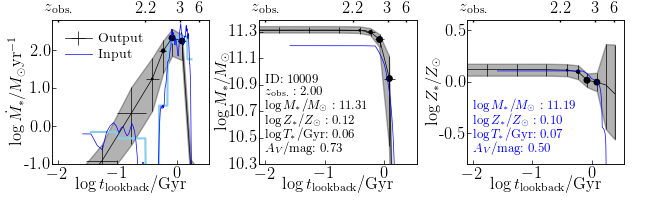}
	\includegraphics[width=0.7\textwidth]{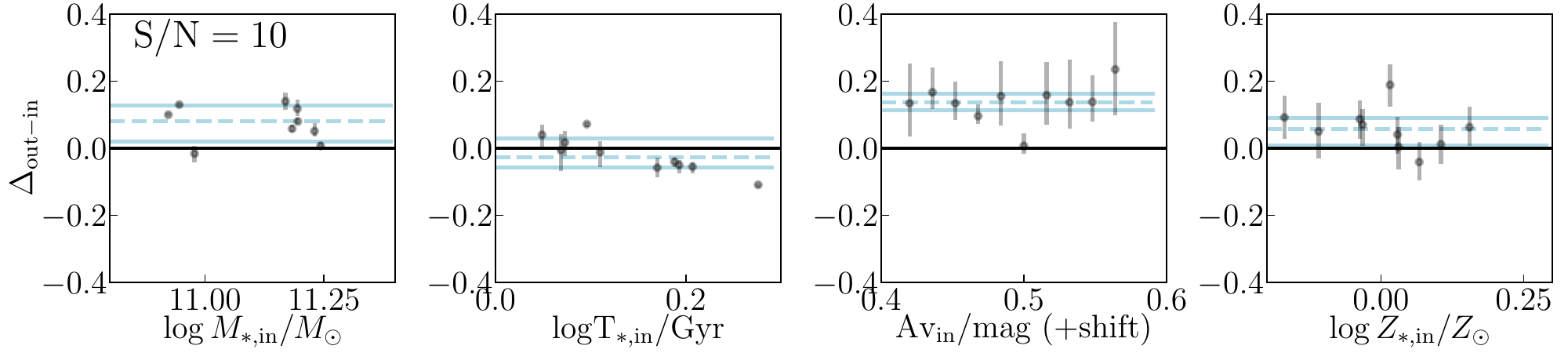}
	\includegraphics[width=0.7\textwidth]{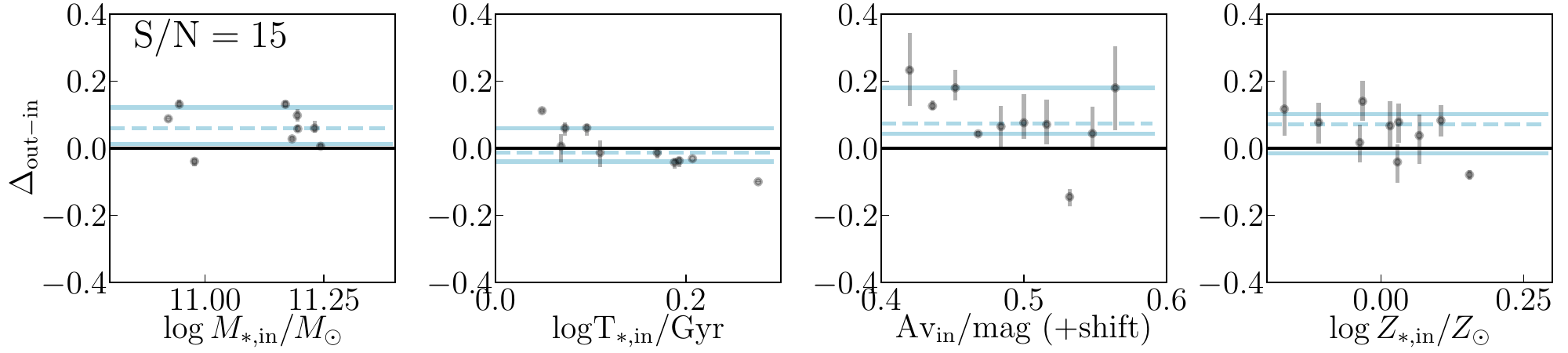}
	\caption{
	({\bf Top}) Results of the star formation reconstruction with \gsf\ for 10 quenching galaxies. The input star formation and metallicity histories (blue lines) are selected from the Illustris simulation within our observed mass range ($\logm\sim11$) and quenched (${\rm SFR}<1\,M_\odot$/yr) at $z\sim2$. The SFHs binned to the fitting template resolution are also shown (light blue) for the comparison with output SFHs. The input and output values for each parameters are shown in the inset (blue and black text, respectively).
	({\bf Bottom}) Summary results of output parameters (${\rm \langle S/N\rangle}=10$ and 15). There is a negative trend in mass-weighted age. This is caused by underestimation of star formation in the oldest bin, but the net effect is small ($\Delta T\sim0.1$\,Gyr). The dust attenuation shows an offset ($\Delta\sim0.2$\,mag), but this is rather dependent on a specific type of SFHs (cf. Figure~\ref{fig:sim}).
	}
\label{fig:sfh_ill}
\end{figure*}

In Figure~\ref{fig:sfh_ill}, we compare the input and output histories for 10 galaxies. The set of galaxies is selected from the Illustris simulation \citep{nelson15}, with a similar mass to our galaxies ($\logm\sim11$) and quenched at the time of observation (SFR$<1\,M_\odot$\,/\,yr at $z\sim2$). The SFHs, and metallicity enrichment histories, are provided to FSPS, to synthesize SEDs. From the generated SEDs, we extract fluxes corresponding to our grism elements (convolved with morphology of one of our sources) and broadband filters. We then add noise with a conservative value of $\langle {\rm S/N}\rangle=10$ at 4200-5000\,\AA (and 15 as a supplemental test). We set the observed redshift to $z=2$ and $A_V=0.5$\,mag uniformly for the sake of simplicity, but this hardly affects the conclusion here. 

In general, the posterior captures the feature of SFHs---the peak time of SFR and its length---, and gives fairly good estimates of SED parameters: stellar mass, mass-weight age and metallicity, and dust attenuation (bottom panel of Figure~\ref{fig:sfh_ill}). There is a trend that \gsf\ underestimates the star formation in the oldest bin. This leads to underestimation in the mass-weight age for old galaxies with $\log T_{*,\rm input}/{\rm Gyr}>0.2$, but this is only for a small amount ($<0.1$\,dex). The offset seen in output metallicity is also small ($\sim0.1$\,dex), and hardly change our conclusion in the main text, as the measurements in the main text quote much larger uncertainties from the analysis in Appendix~A.

One caveat is that dust attenuation is overestimated for $\sim0.2$\,mag (and stellar mass is overestimated for $\sim0.1$\,dex accordingly). However, this should be considered as a result of this specific type of SFHs (and SEDs), as we see fairly good reproduction of the parameter for random SFHs in Figure~\ref{fig:sim}. The offset becomes smaller by increasing the input S/N to 15, but is not completely dismissed.

\begin{deluxetable*}{cccccccccccccc}
\tabletypesize{\footnotesize}
\tablecolumns{14}
\tablewidth{0pt} 
\tablecaption{Summary of physical parameters.}
\tablehead{
\colhead{Obj. ID} & \colhead{R.A.} & \colhead{Decl.} & \colhead{$z_{\rm grism}$} & \colhead{$\log M_*$} & \colhead{$\log Z_*$} & \colhead{$\log T_*$} & \colhead{$A_V$}  & \colhead{$U-V$}  & \colhead{$V-J$} & \multicolumn{2}{c}{S/N$^{\rm a}$} & \colhead{${t_{\rm G102}}^{\rm b}$} & \colhead{${t_{\rm G141}}^{\rm b}$}\\
\colhead{} & \colhead{(deg)} & \colhead{(deg)} & \colhead{} & \colhead{($M_\odot$)} & \colhead{$(Z_\odot)^{\rm c}$} & \colhead{(Gyr)} & \colhead{(mag)}  & \colhead{(mag)} & \colhead{(mag)} & \colhead{Blue} & \colhead{Red} & \colhead{(s)} & \colhead{(s)}
}
\startdata
\cutinhead{MACS J1149.6$+$2223$^{\rm d}$}
00141 & 1.77403e$+02$ & 2.24185e$+01$ & $1.96_{-0.01}^{+0.01}$ & $11.21_{-0.14}^{+0.13}$ & $0.43_{-0.20}^{+0.18}$ & $-0.25_{-0.12}^{+0.11}$ & $1.72_{-0.15}^{+0.20}$ & $1.79_{-0.01}^{+0.01}$ & $1.35_{-0.07}^{+0.10}$ & 4.2 & 24.5 & 9529 & 75987\\
00227 & 1.77407e$+02$ & 2.24162e$+01$ & $2.41_{-0.01}^{+0.01}$ & $10.94_{-0.13}^{+0.15}$ & $-0.28_{-0.17}^{+0.19}$ & $0.23_{-0.14}^{+0.14}$ & $0.74_{-0.08}^{+0.07}$ & $1.20_{-0.01}^{+0.01}$ & $0.81_{-0.05}^{+0.00}$ & 6.8 & 17.4 & 19758 & 80399\\

\cutinhead{GOODS-North}
06215 & 1.89029e$+02$ & 6.21726e$+01$ & $2.30_{-0.02}^{+0.02}$ & $11.55_{-0.13}^{+0.13}$ & $0.36_{-0.17}^{+0.18}$ & $0.01_{-0.09}^{+0.12}$ & $0.65_{-0.09}^{+0.08}$ & $1.43_{-0.01}^{+0.01}$ & $0.83_{-0.01}^{+0.01}$ & 2.8 & 8.5 & 5011 & 6117\\
07276 & 1.89306e$+02$ & 6.21791e$+01$ & $2.50_{-0.01}^{+0.01}$ & $11.65_{-0.14}^{+0.13}$ & $0.14_{-0.20}^{+0.22}$ & $0.20_{-0.10}^{+0.11}$ & $0.32_{-0.08}^{+0.08}$ & $1.50_{-0.01}^{+0.01}$ & $0.77_{-0.01}^{+0.01}$ & 6.1 & 14.1 & 5011 & 5011\\
11470 & 1.89066e$+02$ & 6.21987e$+01$ & $1.67_{-0.01}^{+0.01}$ & $11.07_{-0.12}^{+0.12}$ & $0.33_{-0.17}^{+0.19}$ & $0.08_{-0.11}^{+0.09}$ & $0.13_{-0.09}^{+0.10}$ & $1.17_{-0.01}^{+0.01}$ & $0.50_{-0.01}^{+0.01}$ & 3.0 & 20.1 & 10023 & 8723\\
17599 & 1.89121e$+02$ & 6.22289e$+01$ & $2.15_{-0.01}^{+0.01}$ & $11.16_{-0.15}^{+0.15}$ & $0.23_{-0.24}^{+0.25}$ & $0.17_{-0.10}^{+0.12}$ & $0.41_{-0.10}^{+0.10}$ & $1.77_{-0.17}^{+0.12}$ & $0.86_{-0.03}^{+0.03}$ & 3.5 & 20.6 & 33482 & 39394\\
17735 & 1.89061e$+02$ & 6.22290e$+01$ & $1.84_{-0.01}^{+0.01}$ & $11.33_{-0.13}^{+0.13}$ & $0.36_{-0.17}^{+0.17}$ & $0.14_{-0.16}^{+0.14}$ & $0.50_{-0.10}^{+0.11}$ & $1.46_{-0.01}^{+0.01}$ & $0.83_{-0.00}^{+0.00}$ & 3.8 & 23.5 & 5011 & 8123\\
19341 & 1.89087e$+02$ & 6.22367e$+01$ & $1.86_{-0.01}^{+0.01}$ & $10.98_{-0.16}^{+0.14}$ & $0.02_{-0.26}^{+0.26}$ & $0.29_{-0.09}^{+0.12}$ & $0.04_{-0.04}^{+0.07}$ & $1.19_{-0.01}^{+0.01}$ & $0.57_{-0.01}^{+0.01}$ & 4.1 & 31.1 & 5011 & 38494\\
19850 & 1.89090e$+02$ & 6.22392e$+01$ & $1.86_{-0.01}^{+0.01}$ & $11.06_{-0.16}^{+0.14}$ & $0.25_{-0.20}^{+0.23}$ & $0.15_{-0.14}^{+0.13}$ & $0.53_{-0.15}^{+0.13}$ & $1.59_{-0.01}^{+0.01}$ & $0.90_{-0.01}^{+0.01}$ & 2.1 & 22.2 & 5011 & 38494\\
22774 & 1.89128e$+02$ & 6.22537e$+01$ & $2.01_{-0.01}^{+0.01}$ & $11.37_{-0.12}^{+0.12}$ & $0.14_{-0.19}^{+0.19}$ & $0.16_{-0.10}^{+0.12}$ & $0.78_{-0.13}^{+0.10}$ & $1.64_{-0.01}^{+0.01}$ & $0.97_{-0.01}^{+0.01}$ & 2.2 & 18.4 & 4811 & 31476\\
23006 & 1.89351e$+02$ & 6.22547e$+01$ & $2.48_{-0.01}^{+0.01}$ & $11.11_{-0.14}^{+0.12}$ & $0.11_{-0.20}^{+0.21}$ & $0.18_{-0.10}^{+0.11}$ & $0.05_{-0.05}^{+0.07}$ & $1.27_{-0.01}^{+0.01}$ & $0.55_{-0.01}^{+0.01}$ & 2.8 & 8.3 & 5011 & 4911\\
23249 & 1.89064e$+02$ & 6.22560e$+01$ & $2.37_{-0.01}^{+0.01}$ & $10.89_{-0.14}^{+0.15}$ & $0.10_{-0.28}^{+0.25}$ & $0.21_{-0.27}^{+0.18}$ & $0.34_{-0.22}^{+0.18}$ & $1.28_{-0.01}^{+0.01}$ & $0.56_{-0.03}^{+0.02}$ & 2.9 & 7.7 & 5011 & 15841\\
24033 & 1.89115e$+02$ & 6.22594e$+01$ & $1.67_{-0.01}^{+0.01}$ & $11.21_{-0.15}^{+0.14}$ & $0.47_{-0.19}^{+0.18}$ & $0.19_{-0.10}^{+0.12}$ & $0.25_{-0.08}^{+0.09}$ & $1.54_{-0.01}^{+0.01}$ & $0.87_{-0.00}^{+0.00}$ & 3.2 & 29.2 & 4811 & 39494\\
33780 & 1.89202e$+02$ & 6.23172e$+01$ & $1.87_{-0.01}^{+0.01}$ & $11.71_{-0.19}^{+0.17}$ & $0.48_{-0.20}^{+0.22}$ & $0.30_{-0.09}^{+0.08}$ & $0.51_{-0.12}^{+0.12}$ & $1.92_{-0.01}^{+0.01}$ & $1.31_{-0.01}^{+0.01}$ & 2.9 & 14.8 & 33282 & 14635\\

\cutinhead{GOODS-South}
09704 & 5.32857e$+01$ & -2.78641e$+01$ & $1.74_{-0.01}^{+0.01}$ & $11.71_{-0.14}^{+0.14}$ & $0.40_{-0.18}^{+0.19}$ & $0.28_{-0.12}^{+0.12}$ & $0.71_{-0.21}^{+0.14}$ & $1.69_{-0.02}^{+0.02}$ & $1.14_{-0.00}^{+0.00}$ & 12.5 & 17.4 & 98073 & 4711\\
23073 & 5.31231e$+01$ & -2.78034e$+01$ & $2.34_{-0.01}^{+0.01}$ & $11.32_{-0.18}^{+0.15}$ & $0.34_{-0.21}^{+0.22}$ & $0.10_{-0.10}^{+0.10}$ & $0.15_{-0.09}^{+0.09}$ & $1.46_{-0.01}^{+0.01}$ & $0.68_{-0.01}^{+0.01}$ & 5.4 & 11.9 & 0 & 9423\\
24569 & 5.31588e$+01$ & -2.77972e$+01$ & $1.90_{-0.01}^{+0.01}$ & $11.17_{-0.18}^{+0.16}$ & $-0.14_{-0.23}^{+0.25}$ & $0.35_{-0.09}^{+0.09}$ & $0.20_{-0.08}^{+0.07}$ & $1.55_{-0.01}^{+0.00}$ & $0.77_{-0.01}^{+0.01}$ & 1.0 & 22.6 & 103246 & 23358\\
31397 & 5.31410e$+01$ & -2.77667e$+01$ & $1.91_{-0.01}^{+0.01}$ & $11.51_{-0.21}^{+0.21}$ & $0.45_{-0.25}^{+0.26}$ & $0.15_{-0.10}^{+0.08}$ & $0.38_{-0.08}^{+0.10}$ & $1.66_{-0.00}^{+0.00}$ & $0.94_{-0.00}^{+0.00}$ & 5.9 & 32.3 & 0 & 21552\\
39364 & 5.30628e$+01$ & -2.77265e$+01$ & $1.61_{-0.01}^{+0.01}$ & $11.53_{-0.21}^{+0.23}$ & $0.17_{-0.25}^{+0.23}$ & $0.44_{-0.18}^{+0.12}$ & $0.22_{-0.07}^{+0.08}$ & $1.62_{-0.01}^{+0.01}$ & $1.09_{-0.00}^{+0.00}$ & 8.9 & 23.6 & 27270 & 8923\\
41021 & 5.31874e$+01$ & -2.77192e$+01$ & $2.32_{-0.01}^{+0.01}$ & $11.28_{-0.13}^{+0.12}$ & $0.08_{-0.17}^{+0.16}$ & $0.16_{-0.12}^{+0.10}$ & $0.27_{-0.09}^{+0.09}$ & $1.23_{-0.01}^{+0.01}$ & $0.54_{-0.01}^{+0.01}$ & 4.9 & 15.7 & 0 & 4711\\
41148 & 5.31279e$+01$ & -2.77189e$+01$ & $1.76_{-0.01}^{+0.01}$ & $11.47_{-0.21}^{+0.20}$ & $0.35_{-0.26}^{+0.26}$ & $0.39_{-0.10}^{+0.10}$ & $0.08_{-0.07}^{+0.07}$ & $1.87_{-0.02}^{+0.02}$ & $1.18_{-0.01}^{+0.01}$ & 3.4 & 10.6 & 23058 & 4611\\
41520 & 5.31527e$+01$ & -2.77163e$+01$ & $1.60_{-0.01}^{+0.01}$ & $11.31_{-0.22}^{+0.25}$ & $0.20_{-0.27}^{+0.25}$ & $0.46_{-0.18}^{+0.13}$ & $0.14_{-0.07}^{+0.07}$ & $1.81_{-0.01}^{+0.01}$ & $1.18_{-0.00}^{+0.00}$ & 4.7 & 10.2 & 27470 & 4711\\
43005 & 5.31085e$+01$ & -2.77101e$+01$ & $1.60_{-0.01}^{+0.01}$ & $11.32_{-0.22}^{+0.23}$ & $0.38_{-0.25}^{+0.29}$ & $0.28_{-0.09}^{+0.10}$ & $0.39_{-0.08}^{+0.08}$ & $1.82_{-0.01}^{+0.00}$ & $1.12_{-0.00}^{+0.00}$ & 3.9 & 13.1 & 23058 & 9323\\
43114 & 5.30624e$+01$ & -2.77069e$+01$ & $1.90_{-0.01}^{+0.01}$ & $12.06_{-0.20}^{+0.19}$ & $0.07_{-0.19}^{+0.21}$ & $0.37_{-0.10}^{+0.10}$ & $0.27_{-0.07}^{+0.07}$ & $1.45_{-0.00}^{+0.00}$ & $0.81_{-0.00}^{+0.00}$ & 25.2 & 49.4 & 27270 & 7720\\

\enddata
\tablecomments{
\\
$^{\rm a}$Average S\,/Ns of grism spectral element measured at blue ($3400$--$3800$\,\AA) and red ($4200$--$5000$\,\AA) wavelength ranges.\\
$^{\rm b}$ Total exposure time in G102 and G141 observations.\\
$^{\rm c}Z_\odot=0.0142$ \citep{asplund09}.\\
$^{\rm d}$Stellar masses are corrected for magnifications by the foreground cluster.
}
\label{tab:sed}
\end{deluxetable*}

\begin{deluxetable*}{cccccccccccccc}
\tabletypesize{\footnotesize}
\tablecolumns{14}
\tablewidth{0pt} 
\tablecaption{Summary of physical parameters with different SFHs.}
\tablehead{
 &  \multicolumn{3}{c}{\gsf$^{a}$} & \multicolumn{5}{c}{Exponential Model} &  \multicolumn{5}{c}{Delayed Exponential Model}\\ \cmidrule(lr){2-4} \cmidrule(lr){5-9} \cmidrule(lr){10-14} 
\colhead{Obj. ID} & \colhead{$\chi^2/\nu$} & \colhead{{$\log Z_L$}$^{b}$} & \colhead{{$\log T_L$}$^{b}$} & \colhead{$\chi^2/\nu$} & \colhead{$\log M_*$} & \colhead{$\log Z_L$} & \colhead{$\log T_L$} & \colhead{$A_V$} & \colhead{$\chi^2/\nu$} & \colhead{$\log M_*$} & \colhead{$\log Z_L$} & \colhead{$\log T_L$} & \colhead{$A_V$}\\
 & & \colhead{$(Z_\odot)$} & \colhead{(Gyr)} & & \colhead{$(M_\odot)$} & \colhead{$(Z_\odot)$} & \colhead{(Gyr)} & \colhead{(mag)} & & \colhead{$(M_\odot)$} & \colhead{$(Z_\odot)$} & \colhead{(Gyr)} & \colhead{(mag)}
}
\startdata
00141 & $2.41$ & $0.45_{-0.19}^{+0.17}$ & $-0.34_{-0.13}^{+0.09}$         & $2.43$ & $11.49_{-0.04}^{+0.03}$ & $0.54_{-0.04}^{+0.04}$ & $-0.61_{-0.04}^{+0.04}$ & $1.94_{-0.07}^{+0.06}$         & $2.47$ & $11.46_{-0.02}^{+0.01}$ & $0.55_{-0.04}^{+0.04}$ & $-0.59_{-0.04}^{+0.04}$ & $1.95_{-0.06}^{+0.03}$ \\
00227 & $1.75$ & $-0.17_{-0.10}^{+0.12}$ & $-0.18_{-0.14}^{+0.17}$         & $3.52$ & $11.02_{-0.01}^{+0.09}$ & $-0.30_{-0.14}^{+0.14}$ & $-0.41_{-0.07}^{+0.05}$ & $0.93_{-0.01}^{+0.10}$         & $3.95$ & $10.96_{-0.03}^{+0.01}$ & $-0.28_{-0.15}^{+0.12}$ & $-0.39_{-0.04}^{+0.05}$ & $0.85_{-0.14}^{+0.01}$ \\
06215 & $1.59$ & $0.41_{-0.15}^{+0.14}$ & $-0.10_{-0.09}^{+0.10}$         & $1.77$ & $11.48_{-0.01}^{+0.00}$ & $0.01_{-0.07}^{+0.07}$ & $-0.47_{-0.06}^{+0.08}$ & $1.14_{-0.04}^{+0.01}$         & $1.65$ & $11.49_{-0.04}^{+0.00}$ & $0.05_{-0.10}^{+0.15}$ & $-0.40_{-0.04}^{+0.04}$ & $1.11_{-0.04}^{+0.01}$ \\
07276 & $1.54$ & $0.13_{-0.18}^{+0.21}$ & $0.15_{-0.10}^{+0.11}$         & $2.58$ & $11.57_{-0.00}^{+0.00}$ & $0.00_{-0.07}^{+0.07}$ & $0.10_{-0.03}^{+0.03}$ & $0.23_{-0.01}^{+0.01}$         & $2.50$ & $11.52_{-0.00}^{+0.00}$ & $-0.42_{-0.08}^{+0.08}$ & $0.21_{-0.04}^{+0.04}$ & $0.18_{-0.01}^{+0.01}$ \\
09704 & $1.74$ & $0.40_{-0.16}^{+0.17}$ & $0.15_{-0.12}^{+0.13}$         & $3.34$ & $11.59_{-0.02}^{+0.00}$ & $0.54_{-0.04}^{+0.04}$ & $-0.29_{-0.04}^{+0.06}$ & $1.27_{-0.08}^{+0.01}$         & $2.72$ & $11.58_{-0.00}^{+0.00}$ & $0.55_{-0.04}^{+0.04}$ & $-0.20_{-0.04}^{+0.04}$ & $1.18_{-0.01}^{+0.01}$ \\
11470 & $1.88$ & $0.33_{-0.14}^{+0.14}$ & $-0.11_{-0.11}^{+0.11}$         & $2.65$ & $11.05_{-0.00}^{+0.00}$ & $0.20_{-0.07}^{+0.07}$ & $-0.10_{-0.03}^{+0.03}$ & $0.27_{-0.00}^{+0.00}$         & $5.59$ & $11.03_{-0.00}^{+0.03}$ & $0.50_{-0.14}^{+0.06}$ & $-0.10_{-0.03}^{+0.03}$ & $0.14_{-0.01}^{+0.09}$ \\
17599 & $1.49$ & $0.27_{-0.22}^{+0.21}$ & $0.14_{-0.09}^{+0.10}$         & $3.69$ & $11.13_{-0.08}^{+0.03}$ & $0.03_{-0.10}^{+0.47}$ & $-0.03_{-0.10}^{+0.13}$ & $0.52_{-0.29}^{+0.11}$         & $2.74$ & $11.11_{-0.04}^{+0.10}$ & $0.52_{-0.53}^{+0.06}$ & $-0.07_{-0.05}^{+0.08}$ & $0.50_{-0.17}^{+0.15}$ \\
17735 & $1.70$ & $0.40_{-0.14}^{+0.15}$ & $-0.02_{-0.12}^{+0.13}$         & $6.68$ & $11.28_{-0.00}^{+0.01}$ & $0.04_{-0.10}^{+0.50}$ & $-0.22_{-0.07}^{+0.05}$ & $0.87_{-0.04}^{+0.00}$         & $2.24$ & $11.27_{-0.00}^{+0.00}$ & $0.40_{-0.07}^{+0.07}$ & $-0.20_{-0.04}^{+0.04}$ & $0.81_{-0.00}^{+0.00}$ \\
19341 & $4.37$ & $0.08_{-0.16}^{+0.17}$ & $0.19_{-0.10}^{+0.13}$         & $20.66$ & $10.79_{-0.01}^{+0.00}$ & $-0.40_{-0.07}^{+0.07}$ & $-0.50_{-0.03}^{+0.04}$ & $0.73_{-0.01}^{+0.01}$         & $4.79$ & $10.79_{-0.01}^{+0.00}$ & $-0.41_{-0.07}^{+0.07}$ & $-0.49_{-0.04}^{+0.04}$ & $0.73_{-0.01}^{+0.01}$ \\
19850 & $1.87$ & $0.27_{-0.20}^{+0.22}$ & $0.06_{-0.12}^{+0.13}$         & $4.63$ & $11.05_{-0.00}^{+0.00}$ & $-0.75_{-0.03}^{+0.03}$ & $-0.10_{-0.03}^{+0.03}$ & $1.03_{-0.01}^{+0.01}$         & $2.43$ & $11.03_{-0.03}^{+0.03}$ & $0.26_{-0.17}^{+0.28}$ & $-0.03_{-0.08}^{+0.05}$ & $0.63_{-0.04}^{+0.10}$ \\
22774 & $1.19$ & $0.17_{-0.19}^{+0.18}$ & $0.09_{-0.09}^{+0.10}$         & $2.11$ & $11.36_{-0.00}^{+0.00}$ & $-0.75_{-0.03}^{+0.03}$ & $-0.10_{-0.03}^{+0.03}$ & $1.13_{-0.01}^{+0.01}$         & $1.59$ & $11.33_{-0.00}^{+0.00}$ & $-0.19_{-0.08}^{+0.08}$ & $-0.10_{-0.04}^{+0.04}$ & $0.98_{-0.03}^{+0.03}$ \\
23006 & $1.06$ & $0.18_{-0.17}^{+0.17}$ & $0.08_{-0.09}^{+0.10}$         & $1.52$ & $11.03_{-0.01}^{+0.05}$ & $-0.64_{-0.10}^{+0.09}$ & $-0.00_{-0.04}^{+0.04}$ & $0.19_{-0.05}^{+0.02}$         & $1.34$ & $11.05_{-0.02}^{+0.05}$ & $-0.66_{-0.10}^{+0.11}$ & $0.00_{-0.04}^{+0.03}$ & $0.23_{-0.08}^{+0.06}$ \\
23073 & $1.19$ & $0.35_{-0.20}^{+0.21}$ & $0.07_{-0.10}^{+0.10}$         & $3.62$ & $11.29_{-0.00}^{+0.03}$ & $0.25_{-0.10}^{+0.16}$ & $-0.02_{-0.08}^{+0.05}$ & $0.21_{-0.01}^{+0.10}$         & $2.66$ & $11.30_{-0.00}^{+0.00}$ & $0.41_{-0.08}^{+0.08}$ & $-0.01_{-0.04}^{+0.04}$ & $0.22_{-0.01}^{+0.01}$ \\
23249 & $1.48$ & $0.20_{-0.22}^{+0.20}$ & $0.08_{-0.27}^{+0.22}$         & $1.52$ & $10.81_{-0.01}^{+0.02}$ & $0.41_{-0.12}^{+0.12}$ & $-0.53_{-0.08}^{+0.06}$ & $0.77_{-0.02}^{+0.06}$         & $1.45$ & $10.80_{-0.08}^{+0.05}$ & $0.47_{-0.14}^{+0.10}$ & $-0.50_{-0.11}^{+0.12}$ & $0.77_{-0.13}^{+0.06}$ \\
24033 & $2.18$ & $0.46_{-0.20}^{+0.17}$ & $0.11_{-0.09}^{+0.10}$         & $20.81$ & $11.15_{-0.00}^{+0.00}$ & $0.21_{-0.08}^{+0.08}$ & $-0.01_{-0.04}^{+0.04}$ & $0.48_{-0.01}^{+0.00}$         & $5.73$ & $11.16_{-0.00}^{+0.00}$ & $0.40_{-0.07}^{+0.07}$ & $0.00_{-0.03}^{+0.03}$ & $0.48_{-0.00}^{+0.00}$ \\
24569 & $2.66$ & $-0.08_{-0.22}^{+0.24}$ & $0.32_{-0.09}^{+0.09}$         & $6.03$ & $10.91_{-0.00}^{+0.01}$ & $-0.20_{-0.07}^{+0.07}$ & $0.10_{-0.03}^{+0.03}$ & $0.03_{-0.01}^{+0.03}$         & $5.16$ & $10.91_{-0.01}^{+0.01}$ & $-0.20_{-0.07}^{+0.07}$ & $0.10_{-0.03}^{+0.03}$ & $0.05_{-0.02}^{+0.03}$ \\
31397 & $2.81$ & $0.43_{-0.22}^{+0.23}$ & $0.10_{-0.09}^{+0.09}$         & $8.22$ & $11.48_{-0.04}^{+0.00}$ & $0.36_{-0.14}^{+0.09}$ & $0.02_{-0.05}^{+0.08}$ & $0.52_{-0.16}^{+0.01}$         & $4.89$ & $11.48_{-0.00}^{+0.00}$ & $0.41_{-0.08}^{+0.07}$ & $0.10_{-0.04}^{+0.04}$ & $0.41_{-0.01}^{+0.01}$ \\
33780 & $1.72$ & $0.48_{-0.21}^{+0.19}$ & $0.27_{-0.09}^{+0.08}$         & $2.95$ & $11.55_{-0.00}^{+0.00}$ & $-0.75_{-0.03}^{+0.03}$ & $0.00_{-0.03}^{+0.03}$ & $1.35_{-0.01}^{+0.01}$         & $1.81$ & $11.66_{-0.03}^{+0.03}$ & $-0.27_{-0.16}^{+0.12}$ & $0.35_{-0.07}^{+0.07}$ & $0.67_{-0.10}^{+0.13}$ \\
39364 & $3.98$ & $0.17_{-0.19}^{+0.19}$ & $0.36_{-0.16}^{+0.13}$         & $6.42$ & $11.44_{-0.00}^{+0.00}$ & $-0.61_{-0.09}^{+0.09}$ & $0.00_{-0.03}^{+0.03}$ & $1.19_{-0.05}^{+0.00}$         & $7.11$ & $11.45_{-0.04}^{+0.00}$ & $-0.56_{-0.09}^{+0.13}$ & $0.00_{-0.04}^{+0.04}$ & $1.24_{-0.06}^{+0.00}$ \\
41021 & $2.86$ & $0.20_{-0.15}^{+0.15}$ & $0.03_{-0.13}^{+0.11}$         & $3.22$ & $11.19_{-0.12}^{+0.01}$ & $0.13_{-0.86}^{+0.15}$ & $-0.47_{-0.10}^{+0.38}$ & $0.71_{-0.31}^{+0.05}$         & $3.00$ & $11.29_{-0.00}^{+0.03}$ & $-0.72_{-0.06}^{+0.95}$ & $-0.13_{-0.38}^{+0.06}$ & $0.41_{-0.01}^{+0.32}$ \\
41148 & $2.36$ & $0.36_{-0.25}^{+0.26}$ & $0.39_{-0.10}^{+0.10}$         & $3.65$ & $11.56_{-0.00}^{+0.00}$ & $0.20_{-0.07}^{+0.07}$ & $0.48_{-0.02}^{+0.02}$ & $0.00_{-0.00}^{+0.00}$         & $3.77$ & $11.55_{-0.00}^{+0.00}$ & $0.20_{-0.07}^{+0.07}$ & $0.47_{-0.02}^{+0.02}$ & $0.00_{-0.00}^{+0.00}$ \\
41520 & $1.94$ & $0.19_{-0.27}^{+0.24}$ & $0.45_{-0.18}^{+0.12}$         & $2.88$ & $11.36_{-0.00}^{+0.00}$ & $0.20_{-0.07}^{+0.07}$ & $0.48_{-0.02}^{+0.02}$ & $0.06_{-0.01}^{+0.01}$         & $2.71$ & $11.35_{-0.00}^{+0.01}$ & $0.20_{-0.07}^{+0.07}$ & $0.47_{-0.02}^{+0.02}$ & $0.06_{-0.01}^{+0.01}$ \\
43005 & $2.73$ & $0.38_{-0.26}^{+0.29}$ & $0.28_{-0.09}^{+0.10}$         & $3.27$ & $11.20_{-0.00}^{+0.00}$ & $0.55_{-0.03}^{+0.03}$ & $0.20_{-0.03}^{+0.03}$ & $0.05_{-0.01}^{+0.01}$         & $4.81$ & $11.21_{-0.00}^{+0.00}$ & $0.55_{-0.03}^{+0.03}$ & $0.20_{-0.03}^{+0.03}$ & $0.05_{-0.01}^{+0.01}$ \\
43114 & $8.08$ & $0.24_{-0.16}^{+0.16}$ & $0.24_{-0.12}^{+0.12}$         & $12.98$ & $11.99_{-0.00}^{+0.02}$ & $-0.42_{-0.13}^{+0.08}$ & $-0.10_{-0.04}^{+0.03}$ & $0.81_{-0.00}^{+0.06}$         & $17.93$ & $11.95_{-0.01}^{+0.03}$ & $-0.55_{-0.10}^{+0.16}$ & $-0.17_{-0.05}^{+0.08}$ & $0.92_{-0.01}^{+0.04}$ \\

\enddata
\tablecomments{
\\
$^{\rm a}$Stellar mass and dust attenuation for \gsf\ are listed in Table~\ref{tab:sed}.\\
$^{\rm b}$Light-weighted metallicity and age, to match those derived with functional SFHs.\\
}
\label{tab:SFH2}
\end{deluxetable*}

\clearpage

\bibliographystyle{apj}
\bibliography{./ms.bbl}
\end{document}